\documentclass[12pt]{article}
\usepackage[dvips]{graphicx}
\def\be{\begin{equation}}
\def\ee{\end{equation}}
\def\ba{\begin{eqnarray}}
\def\ea{\end{eqnarray}}

\newcommand{\mc}{\multicolumn}
\newcommand{\CC}{\phantom{1}}
\newcommand{\mx}{\phantom{1}}
\newcommand{\mxx}{\phantom{11}}
\newcommand{\BT}{\begin{table}\small\begin{center}\begin{tabular}}
\newcommand{\ET}{\end{center} \end{table}}
\newcommand{\PB}{\parbox[t]{.85\textwidth}}
\newcommand{\LL}{\langle}
\newcommand{\GG}{\rangle}

\begin{document}
\begin{titlepage}
\thispagestyle{empty}
\vskip0.5cm
\begin{flushright}
HUB-EP-98/31  \\
MS--TPI--98--10
\end{flushright}
\vskip0.8cm
\begin{center}
{\Large {\bf Critical Exponents of the 3D Ising}}
\vskip3mm
{\Large {\bf Universality Class From Finite Size Scaling}}
\vskip3mm
{\Large {\bf With Standard and Improved Actions}}
\end{center}
\vskip1.5cm
\begin{center}
{\large M. Hasenbusch${}^a$, K. Pinn${}^b$, and S. Vinti${}^b$}\\
\vskip5mm
${}^a$ Fachbereich Physik, Humboldt-Universit\"at zu Berlin\\ 
Invalidenstr.\ 110, D--10099 Berlin, Germany \\
e--mail: {\sl hasenbus@ficus1.physik.hu-berlin.de}
\vskip5mm
${}^b$ Institut f\"ur Theoretische Physik I, Universit\"at M\"unster\\ 
Wilhelm--Klemm--Str.~9, D--48149 M\"unster, Germany \\
e--mail: {\sl pinn@uni--muenster.de, vinti@uni--muenster.de}
\end{center}
\vskip1.0cm
\begin{abstract}
\par\noindent
We propose a method to obtain an improved Hamiltonian (action) for the
Ising universality class in three dimensions. The improved Hamiltonian
has suppressed leading corrections to scaling. It is obtained by
tuning models with two coupling constants.  We studied three different
models: the $\pm 1$ Ising model with nearest neighbour and body
diagonal interaction, the spin-1 model with states $0,\pm 1$, and
nearest neighbour interaction, and $\phi^4$-theory on the lattice
(Landau-Ginzburg Hamiltonian).  The remarkable finite size scaling
properties of the suitably tuned spin--1 model are compared in detail
with those of the standard Ising model. Great care is taken to
estimate the systematic errors from residual corrections to scaling.
Our best estimates for the critical exponents are $\nu= 0.6298(5)$ and
$\eta= 0.0366(8)$, where the given error estimates take into account the
statistical and systematic uncertainties.
\end{abstract}
\end{titlepage}

\tableofcontents
\newpage 
\listoftables 
\vspace{1cm}
\noindent IA: Improved Action \\
\noindent SA: Standard Action 
\newpage 
\listoffigures
\vspace{1cm}
\noindent IA: Improved Action \\
\noindent SA: Standard Action 
\newpage

\section{Introduction}

In Monte Carlo simulations the size of the systems that can be studied
is limited by the memory of the computer and by the CPU-time that is
available.  Therefore, in many instances, finite size scaling
\cite{privman} is the key to a precise determination of properties of
statistical systems at criticality.  Finite size scaling laws are
affected by corrections to scaling.  These corrections to scaling
cause systematic errors in the results for the infinite volume limit
one is interested in.  With improving statistical accuracy of the
Monte Carlo data it becomes important to deal properly with systematic
errors.  One way to proceed is to include corrections to scaling into
the fit ans\"atze when analysing the data. Another, more fundamental 
way is to remove corrections already from the system to be studied.

Renormalization group (RG) \cite{RG} offers (at least in principle) a
way to achieve this goal.  
RG fixed point actions\footnote{What is called Hamiltonian or energy
function in Statistical Mechanics is called Action in Euclidean
Field Theory.} are free of
corrections to scaling. However such actions in general contain an
infinite number of couplings.  In practical applications one is forced
to truncate the action to a finite number of terms, which in fact is
an uncontrolled approximation.  For an application of this strategy to
asymptotically free models see the work on 
Perfect Actions~\cite{PerfectAction}.

A different approach was pioneered by K. Symanzik \cite{Symanzik}.
Higher order terms are added to the action.  By imposing certain
conditions on observables leading corrections to scaling are
eliminated. While Symanzik formulated his method in the framework of
perturbation theory, recently there have been attempts to apply this
method in a non-perturbative, i.e. numerical, setting
\cite{cloverterm}.

Our present approach is closer to this latter point of view than to
the block spin renormalization group inspired framework.

The idea to improve the scaling properties of Ising models by moving
to models with generalized actions and tuning the coupling constants
as to obtain reduced corrections to scaling was already followed in
the work of Bl\"ote et al.~\cite{bloete}.  However, in their work they
do not uncover the principle and method they used to obtain their
improved actions.

In a recent paper \cite{letter}, we presented data obtained with a
spin-1 Ising model with states $0,\pm 1$. Tuning the two coupling
constants in the proper way, we were able to reduce the corrections to
scaling in various quantities dramatically. Especially the Binder
cumulant and its derivatives, and also the susceptibility, could be
fitted with scaling laws without corrections to scaling terms,
yielding very precise estimates of the 3D Ising critical exponents.

Recently the authors of refs.~\cite{ball1,ball2} argued that the
improvement discussed above does not lead to reduced error estimates
for critical exponents. They argue that the error estimates given in
our recent paper \cite{letter} are underestimated, since we do not
take into account residual leading correction to scaling corrections.
Such corrections might well be present, since the parameters of our
improved model are computed numerically. However, this is not
the full story as we shall explain in this paper. Our argument is
based on the fact that ratios of correction to scaling amplitudes are
universal.

In this article, we describe in detail our method, the numerical
results, and the fitting procedures. We confront the results from the
improved actions with high precision data from simulations of the
standard Ising model, estimating with a well defined procedure
systematic errors for both actions.

\section{Improving the Scaling Behaviour}

\subsection{The Models}

Usually, Monte Carlo studies of the Ising model are done 
using what in field theory is called 

\vskip5mm
\noindent{\bf Standard Action}
\be
S = - \beta \sum_{<i,j>} s_i \, s_j \, .
\ee 
The $s_i$ take values $\pm 1$, and the spin-spin interaction is a sum
over all nearest neighbour pairs $<i,j>$.  A precise estimate for the
critical coupling was obtained in reference \cite{talapov}:
$\beta_c= 0.2216544(3)(3)$.

In the following we will introduce and study three different models in
the 3D Ising universality class, each of them governed by {\it two}
coupling constants. In all three cases the Boltzmann factor is given
by $\exp(-S)$.

\vskip5mm
\noindent{\bf Spin-1 Model}
\be
S = - \beta \sum_{<i,j>} s_i \, s_j + D \sum_{i} s_i^2 \, .
\ee 
The $s_i$ take values $0,\pm 1$, and the spin-spin interaction
is a sum over all nearest neighbour pairs. This model was 
introduced and studied in \cite{bloete}. There, $D$ was fixed 
to $\ln 2$. The critical $\beta$
corresponding to this particular value of $D$ was estimated in 
\cite{bloete} to be $\beta_c = 0.3934224(10)$.

\vskip5mm
\noindent{\bf NNN Model}
\be
S = - \beta_1 \sum_{<i,j>} s_i \, s_j - \beta_2
\sum_{[i,j]} s_i \, s_j \, .
\ee 
The $s_i$ take values $\pm 1$, and the spin-spin interaction is a sum
over all nearest neighbour pairs $<i,j>$ and third neighbour pairs
(body diagonals) $[i,j]$. Bl\"ote et al.\ fixed $\beta_2/\beta_1= 0.4$
and obtained $\beta_{1,c}= 0.1280036(5)$.

\vskip5mm
\noindent{\bf $\phi^4$-Model}
\be 
S = - \beta \sum_{<i,j>} \phi_i \phi_j 
+ \sum_i \phi_i^2 + \lambda \sum_i (\phi_i^2 - 1)^2  \, .
\ee 
The variables $\phi$ assume real values. In the limit 
$\lambda \rightarrow \infty$ one recovers the standard 
Ising model.

The three models defined above have in common that they have a second
order critical line in the space spanned by the two coupling
constants. We exploit the degree of freedom of moving on the critical
line to find critical models with reduced corrections to scaling.  How
this is done will be explained in the following subsections.

\subsection{Matching of Phenomenological Couplings}

We study two independent phenomenological couplings of the 3D
Ising model, to be called $R_i$, $i=1,2$ in the following.  Both
quantities are universal, i.e.\ at criticality their infinite volume
limit $R_i^*$ does not depend on details of the microscopic
Hamiltonian.  $R_1$ is the ratio of partition functions with
anti-periodic and periodic boundary conditions, respectively,
\be\label{defR1}
R_1 = Z_a / Z_p \, .  
\ee
The lattices will always be cubical with extension $L$ in each of the
three directions. Antiperiodic boundary conditions are imposed only in
one of the three lattice directions.
$R_2$ is the Binder cumulant, 
\be\label{defR2}
R_2 = Q= \frac{\langle m^2 \rangle^2}{\langle m^4 \rangle} \, .
\ee   
Here, $m$ denotes the magnetization per spin, 
\be
\label{defmag} 
m = L^{-3} \sum_i s_i \, . 
\ee 
The $R_i$ are $L$-dependent and, of course, functions of the coupling
parameters in the action.  For the two-coupling models defined above,
we define ``flows'' (lines of constant physics)
$(K_1(L),K_2(L))$ by requiring that
\be
\label{matcheq}
 R_i(L,K_1(L),K_2(L))= R_i^* \, . 
\ee  
$K_1$ and $K_2$ represent the two coupling constants of the model.  In
the next subsection we shall demonstrate that with increasing $L$ the
flows of $(K_1,K_2)$ converge towards a critical point which has
no leading order corrections to scaling.

\subsection{RG Analysis of the Matching Condition}
\label{RGana}

The main features of the two-coupling models can be discussed in the
framework of the renormalization group.  The scaling properties
can be derived from the linearized RG transformation at the fixed
point.

We consider general Hamiltonians with couplings $K_{\alpha}$, where
$\alpha=1,2,\dots$  An RG transformation, realized,
e.g., by a block spin transformation, changes these couplings
according to
\be
K \rightarrow  K' = R(K) \equiv K'(K) \, . 
\ee
A fixed point $K^*$ is defined through $R(K^*)=K^*$.
The linearized transformation at the fixed point can be 
represented by a matrix 
\be
T_{\alpha \beta} = \left. 
\frac{\partial K'_\alpha}{\partial K_\beta} \right|_{K=K^*} \, .
\ee
One introduces ``normal coordinates'' (scaling fields) by 
\be
u_i = u_i(K) = \sum_{\alpha} \varphi_{i,\alpha} \, 
(K_\alpha - K_\alpha^*) \, , 
\ee 
where $\varphi_i$ denotes the $i$-th (left) eigenvector of the matrix $T$, 
$$
\sum_{\alpha} \varphi_{i,\alpha} T_{\alpha\beta} = \lambda_i \, 
\varphi_{i,\beta} \, . 
$$
The $u_i$ transform under RG transformations like
$$
u_i \rightarrow \lambda_i u_i \, . 
$$
In Ising type models the leading eigenvalues are given by 
\be
 \lambda_1 = b^{1/\nu} \, , \;\;\;
 \lambda_2 = b^{-\omega} \, , \;\;\;
 \lambda_3 = b^{-x} \, ,
\ee
where $x> \! > \omega$. Note that $x=2$ for the lattice Gaussian
model. From leading order $\epsilon$-expansion one expects that $x$ is
close to 2 at the Wilson-Fisher fixed point.  $b$ denotes the scale
factor of the RG transformation.

Let us now assume that we have only two non-vanishing couplings
$K_1$ and $K_2$ in our Hamiltonian. Let us then write down explicitly the
condition for  being critical ($u_1=0$) and eliminating the leading
corrections to scaling ($u_2=0$). The first condition reads 
\be
  \varphi_{1,1} (K_1 - K_1^*) 
+ \varphi_{1,2} (K_2 - K_2^*) = \kappa_{1,3} \, , 
\ee
whereas the condition $u_2=0$ translates to 
\be
\label{u2eq}
  \varphi_{2,1} (K_1 - K_1^*) 
+ \varphi_{2,2} (K_2 - K_2^*) = \kappa_{2,3} \, . 
\ee
For $i=1,2$, the $\kappa_{i,3}$ are given by 
\be
\kappa_{i,3} = \sum_{\alpha \geq 3} \varphi_{i,\alpha}
K_{\alpha}^* \, . 
\ee

Let us now study how our matching procedure with the two quantities
$R_1 = Z_a/Z_p$ and $R_2= Q$ works.  The $R_k$ are functions of the
bare couplings and the lattice size:
\be
R_k=R_k(L,K_1,K_2) \, . 
\ee 
We express these quantities as functions of the scaling fields
defined above, 
\be
 R_k(L,K_1,K_2)=
 R_k \left( L^{1/\nu} u_1^{(1)}, L^{-\omega} u_2^{(1)} \right) \, .
\ee
Here, the upper index ${}^{(1)}$ indicates that the scaling field is taken
at the scale of the lattice spacing. The prefactor promotes the
scaling field to its value at the scale $L$.  Taylor-expansion of the
$R_k$ around their fixed point values yields 
\be 
R_k \approx R_k^* + r_{k,1} L^{1/\nu} u_1^{(1)}
                  + r_{k,2} L^{-\omega} u_2^{(1)} \, . 
\ee
The matching conditions $R_k=R_k^*$ are thus equivalent to 
\be
r_{k,1} L^{1/\nu} u_1^{(1)} + r_{k,2} L^{-\omega} u_2^{(1)} = 0 \, ,
\ee
for $k=1,2$.  We obtain, as desired, the solution $u_1^{(1)}=0$ 
(criticality) and $u_2^{(1)} =0$ (no leading order
corrections).  Including higher order corrections in the scaling ansatz,
governed by $u_3$ and exponent $\lambda_3$, one can convince oneself
that fixing $R_1$ and $R_2$ to their fixed point values leads to
convergence to the critical line $u_1=0$ with corrections that decay
like $L^{-x-1/\nu}$. The $u_2=0$ condition is approached with a
much slower rate, namely like $L^{-x+\omega}$.

\subsection{Computing the Matching Flows}

For the three 2-coupling models specified above, we set up a procedure
to determine the flows of couplings $(K_1(L),K_2(L))$ such that
eq.~(\ref{matcheq}) was fulfilled. To this end we used estimates for
$R_1^*=0.5425(10)$ and $R_2^*= 0.6240(10)$, which were an outcome of a
preliminary analysis of data obtained using the standard action.

The matching couplings were searched for using a Newton iteration,
based on the inversion of a matrix made up from Monte Carlo estimates
of the derivatives of the $R_i$ with respect to the two couplings.
Typically three to four iterations were sufficient to find couplings
such that $Z_a/Z_p$ and $Q$ attained the prescribed values within the
given statistical precision.  The results are given in
table~\ref{flows}.

\BT{rllll}
\hline
\hline 
\mc{5}{c}{} \\[-3mm]
\mc{5}{c}{\bf Spin-1 Model} \\[1mm]
   $L$  &  $\beta$ &  $D$ &  $Z_a/Z_p$ & $Q$    \\
\hline
 3 &  0.35737  & 0.4401   & 0.54201(30) & 0.62447(21) \\
 4 &  0.37250  & 0.5510   & 0.54242(25) & 0.62347(18) \\
 5 &  0.37794  & 0.5883   & 0.54292(19) & 0.62421(15) \\
 6 &  0.38210  & 0.6169   & 0.54221(49) & 0.62426(35) \\
 7 &  0.38419  & 0.6311   & 0.54326(30) & 0.62366(22) \\
 8 &  0.38320  & 0.6241   & 0.54291(37) & 0.62430(27) \\
 9 &  0.38320  & 0.6241   & 0.54259(60) & 0.62403(44) \\
10 &  0.38320  & 0.6241   & 0.54288(57) & 0.62431(42) \\
\hline 
\hline 
\mc{5}{c}{} \\[-3mm]
\multicolumn{5}{c}{\bf NNN Model} \\[1mm]
   $L$  & $\beta_1$ & $\beta_2$ & $Z_a/Z_p$ & $Q$    \\
\hline 
 4 &  0.12266 & 0.05406 & 0.5429(2) &  0.6238(2) \\
 5 &  0.12928 & 0.05028 & 0.5427(1) &  0.6241(2) \\
 6 &  0.13431 & 0.04734 & 0.5425(1) &  0.62441(8) \\
 7 &  0.13800 & 0.04518 & 0.5425(3) &  0.6242(2) \\
 8 &  0.14069 & 0.04361 & 0.5426(2) &  0.6243(1) \\
 9 &  0.14292 & 0.04231 & 0.5430(2) &  0.6232(1) \\
10 &  0.14406 & 0.04165 & 0.5425(2) &  0.6242(2) \\
11 &  0.14590 & 0.04059 & 0.5429(1) &  0.62385(8) \\
12 &  0.14724 & 0.03981 & 0.5426(1) &  0.6235(1) \\
13 &  0.14808 & 0.03933 & 0.5431(1) &  0.6236(1) \\
\hline
\hline 
\mc{5}{c}{} \\[-3mm]
\multicolumn{5}{c}{\bf $\phi^4$-Model} \\[1mm]
   $L$  & $\beta$ & $\lambda$ & $Z_a/Z_p$ & $Q$    \\
\hline 
 3 &  0.35303 & 1.5248 & 0.5420(3)&  0.6244(2) \\
 4 &  0.36338 & 1.3282 & 0.5425(3)&  0.6243(2) \\
 5 &  0.36908 & 1.2188 & 0.5425(2)&  0.6241(1) \\
 6 &  0.37165 & 1.1689 & 0.5427(2)&  0.6241(1) \\
 7 &  0.37270 & 1.1481 & 0.5424(2)&  0.6243(1) \\
 8 &  0.37308 & 1.1410 & 0.5421(1)&  0.6245(1) \\
 9 &  0.37273 & 1.1479 & 0.5416(3)&  0.6247(2) \\
\hline
\hline 
 \end{tabular}
\parbox[t]{.85\textwidth}
 {
 \caption[Flows of couplings from matching condition]
 {\label{flows} \small
   Flows of couplings defined such that for all lattice sizes the two
   quantities $R_1 = Z_a/Z_p$ and $R_2= Q$ match (to the given
   statistical precision) with their fixed point values
   $R_1^*=0.5425(10)$ and $R_2^*= 0.6240(10)$.
}
}
\end{center}
\end{table}

A first look at the table reveals that both for the spin-1 and the
$\phi^4$-model, the flow converges to a fixed point quickly, whereas
it keeps moving strongly in the case of the NNN model. A rough
explanation of this could be the following: In order to remove the
leading corrections to scaling, one has to move efficiently between
the Gaussian model (equivalent to the $\phi^4$-model with vanishing
$\lambda$) and the non-trivial Wilson-Fisher fixed point. Moving
between these two fixed points is most efficiently done using a
$\phi^4$-type coupling, which is implicitly present also in the spin-1
model. The NNN model seems to need renormalization to a larger scale in
order to come close to the flow line connecting the Gaussian with
the Wilson-Fisher fixed point.

Plotting the second coupling vs.\ the first one, one finds that with
very good precision the critical line can be approximated by a
straight line:
\be
K_2(L)= a_1 + a_2 \, K_1(L) \, , 
\ee
with 
\begin{center}
\begin{tabular}{cll}
model & \phantom{--}$a_1$ & \phantom{--}$a_2$  \\
\hline 
spin-1   &         --2.04     & \phantom{--}6.95   \\  
NNN      & \phantom{--}0.1253    & --0.5804 \\
$\phi^4$ & \phantom{--}8.29      & --19.17   \\
\hline 
\end{tabular}
\end{center}
\vskip8mm
It is assuring that for the spin-1 and the NNN model,
the critical coupling estimates determined by Bl\"ote et al.\
are in good agreement with our critical lines. 

Fits of $K_1(L)$ with a power law 
\be
K_1(L)= c_1 + c_2 \, L^{-\alpha} 
\ee
yielded good fits in all cases, with exponents of order 3 in the cases
of spin-1 and $\phi^4$-models.  Note that this exponent is much larger
than $-x + \omega$ that is to be expected from the theory presented
above. For the NNN model, the exponent is of order 1.

The slow convergence of the NNN flow motivated our decision to discard
this model from further investigation. Also, the behaviour of the
$\phi^4$ and spin-1 models appears to be very similar.  We decided to
concentrate on the spin-1 model, and leave the $\phi^4$-model for
later study.

Note that our result for the optimal $\lambda \approx 1.145$ of the 
$\phi^4$-model is consistent with the observation of ref. \cite{ball2}
that the optimal $\lambda$ should be close to one.

\subsection{Identification of the $u_2=0$ Line}

After the decision to concentrate on the spin-1 model, we determined
an approximation of the $u_2 = 0$ manifold.  This was done by looking
at the derivatives of the $R_i$ at criticality.  From the RG analysis
of section~\ref{RGana} one infers
\be
\frac{\partial R_k}{\partial K_{\alpha}} = 
\sum_i \frac{\partial R_k}{\partial u_i} 
       \frac{\partial u_i}{\partial K_{\alpha}}
=  r_{k,1} L^{1/\nu} \varphi_{1,\alpha}
+  r_{k,2} L^{-\omega} \varphi_{2,\alpha} \, . 
\ee
The left hand side of the equation can be determined by Monte Carlo
simulation and will therefore be assumed as known.  Taking into
account that without loss of generality one can set $\varphi_{i,1}
\equiv 1$, the equations can be solved (or fitted) if the left hand
side is known at least for two different lattice sizes.  We performed
Monte Carlo simulations at $(\beta,D)= (0.3832,0.6241)$ for lattice
size 8, 9, 10, and 11.  Fixing the exponents $\nu = 0.63$ and $\omega=
0.81$ \cite{guida} we obtained that the scaling field $\phi_{2,2}$
should be approximately $-1/3$. Plugging this into the
eq.~(\ref{u2eq}),
\be
  \varphi_{2,1} (K_1 - K_1^*) 
+ \varphi_{2,2} (K_2 - K_2^*) = \kappa_{2,3} \, , 
\ee
one obtains that $3 \beta - D$ should be kept constant.  We used our
simulation point (0.3832,0.6241) to fix this constant. In conclusion,
one should approach criticality by varying $\beta$ while adjusting $D$
according to
\be
\label{goodline}
 D = 3 \, (\beta-0.3832) + 0.6241  \, .
\ee
Of course, a precise estimate of $\beta_c$ along 
this line still has to be determined. This will 
be discussed in the next section. 

The fit result for $\varphi_{1,2}$ can be used to perform a
consistency check.  In terms of the scaling fields, the critical line
is characterized by
\be
 (\beta-\beta^*) +  \varphi_{1,2} (D-D^*)  = 0 \, , 
\ee
ignoring all higher couplings.  Solving for $D$ and plugging in the
``experimental values'' $\beta^*=0.3832$, $ D^*=0.6241$ and
$\varphi_{1,2}=-0.1439$ one obtains with good precision the critical
line approximation $D= -2.04 + 6.95 \, \beta$.
Let us close this section by remarking that errors in the precise estimation
of the $u_2=0$ line affect results for the critical exponents
only weakly, e.g., the effect on the exponent $\nu$ is of  
order $L^{-1/\nu - \omega}$. 

\section{Simulation Parameters and Statistics}

\subsection{Standard Action}

Monte Carlo simulations of the 3D Ising model with standard 
action were performed at $\beta=0.2216545$, which is a good
approximation of the critical coupling \cite{bloete,talapov}.

For cubical lattices of size $L=2$ up to $L=19$ we performed
simulations with the multi-spin demon update \cite{creutz,multidemon}.
The update algorithm is local. Hence one expects a dynamical critical
exponent $z\approx 2$.  However, due to the multi-spin coding
implementation a single sweep can be done substantially faster than
with the cluster algorithm.  Therefore one expects a better
performance for the demon update on such small lattices. 

For a subset of the smaller lattices, and for bigger lattices up to
size 128, cluster update was performed, using a new variant of the
algorithm, the {\em wall cluster algorithm}.  Note that there is quite
a lot of freedom in the selection of clusters which are flipped during
one update step.  In the wall cluster algorithm, one flips all
clusters that intersect with a given lattice plane.
Sequentially one takes lattice planes in $1-2$ , $1-3$ and $2-3$
direction.  The position is chosen randomly.  Let us call the
procedure to generate and flip all the clusters connected to the
selected plane a wall cluster update step.  The motivation for
choosing this type of update was that the construction of all clusters
that have elements in a lattice plane is needed for the measurement of
$Z_a/Z_p$ anyway.  Testing the performance of the algorithm shows that
there is some small gain in efficiency compared to the single cluster
update \cite{papermartin}. Results for cluster size distribution and
relaxation times of simulations at $\beta_c$ are given in
table~\ref{wallcluster}.

\BT{r|c|c|c|c|c}
 $L$ & $M$ & $S/V$ & $\tau_E$ & $\tau_{\chi}$ & $\tau_b$ \\
  \hline
  6 & 300000& 0.38412(29) & 1.12(2) & 1.18(2) & 0.53(1) \\
  8 & 300000& 0.33222(27) & 1.13(2) & 1.19(2) & 0.56(1) \\
 12 & 300000& 0.27005(24) & 1.13(2) & 1.18(2) & 0.57(1) \\
 16 & 300000& 0.23277(23) & 1.17(2) & 1.19(2) & 0.58(1) \\
 24 & 300000& 0.18866(20) & 1.15(2) & 1.13(2) & 0.55(1) \\
 32 & 700000& 0.16214(13) & 1.18(2) & 1.14(2) & 0.56(1) \\
 48 & 450000& 0.13098(14) & 1.20(2) & 1.09(2) & 0.54(1) \\
 64 & 600000& 0.11258(11) & 1.26(2) & 1.10(2) & 0.55(1) \\
 96 & 340000& 0.09075(13) & 1.26(4) & 1.05(3) & 0.53(1) \\
\hline
 \end{tabular}
\PB
 {
 \caption[Testing the wall cluster algorithm]
 {\label{wallcluster}
Testing the performance of the wall cluster algorithm.
$M$ denotes the number of measurements, separated by 
a single wall cluster step. The $\tau$'s 
are integrated autocorrelation times.
The unit of time is set to one complete update of the lattice; 
i.e. one measurement/($S/V$).
$S/V$ denotes the average 
sum of the sizes of the flipped clusters, 
normalized to the lattice volume $L^3$.
}}
\ET
The average sum of the sizes of the clusters per volume that are
flipped in one step behaves as
\be
  S/V = C L^{x} \, . 
\ee 
Fitting the data of table~\ref{wallcluster} to this law, 
discarding the results from $L < 24$ yields 
$ S/V = 1.008(4) L^{-0.527(1)}$ ($\chi^2$/dof=0.4).
The integrated autocorrelation times of the energy, the susceptibility, 
and of the $Z_a/Z_p$ measurements (see below) 
were also fitted to power laws, 
$\tau = c \, L^{z}$, 
using data from all lattice sizes. Only for $\tau_b$ 
the $L=6$ data were discarded. The fit results are 
summarized in the following table: 

\begin{center}
\begin{tabular}{l|c|l|c}
   &   $c$     & \phantom{xx} $z$  &  $\chi^2/$dof \\
\hline 
$\tau_E$       &  1.04(2)  &\phantom{--}0.035(7)    &  0.90 \\
$\tau_{\chi}$  &  1.30(3)  &          --0.044(7)    &  0.57 \\
$\tau_b$       &  0.60(2)  &          --0.028(8)    &  0.71  \\
\hline
\end{tabular}
\end{center}
\vskip1cm

These numbers should  be compared with the corresponding ones 
from the single cluster algorithm \cite{wolff}: 

\begin{center}
\begin{tabular}{r|c|c}
 $ L$ &   $\tau_E$ & $\tau_\chi$ \\
\hline
  16 &  1.36(2) &  1.01(2) \\
  24 &  1.50(3) &  1.06(2) \\
  32 &  1.72(4) &  1.14(3) \\
  48 &  1.90(6) &  1.20(4) \\
  64 &  1.97(5) &  1.20(3) \\
\hline 
\end{tabular}
\end{center}

These $\tau$'s are fitted by $\tau_E \propto L^{0.28(2)}$ and
$\tau_\chi \propto L^{0.14(2)}$. We conclude that the exponents $z$ of
the wall algorithm are smaller than those of the single cluster
algorithm.  For the lattice sizes in question, however, the actual
$\tau$'s are of similar size.

For the measurements of $Z_a/Z_p$ we employed a variant of the
boundary flip algorithm \cite{flip}, where the cluster that has been
built is not flipped but used to construct an observable.

For details on the statistics and lattice sizes of all the runs, see
table~\ref{Statstandard}. The total CPU consumption of the standard
action runs was about 1.1 years on a 200 MHz Pentium Pro PC for the
wall cluster simulations.  All simulations with the demon algorithm
for the standard Ising model took about half a year on a 200 MHz
Pentium Pro PC.

\BT{|rrr|rc|rc|}
\hline 
\mc{5}{|c}{\bf Standard Action} & 
\mc{2}{|c|}{\bf Improved Action} \\
\hline
\mc{3}{|c|}{Cluster} & 
\mc{2}{|c}{Multispin} & \mc{2}{|c|}{Cluster} \\
   $L$  &  stat/3000 & $k$ & $L$ & stat/$32 \cdot 10^6$ & $L$ & stat/$10^6$ \\
\hline
    4   & 100000  &    4   &   2 & 200 &  4& 30 \\
    6   & 100000  &    4   &   3 & 200 &  6& 60 \\
    8   & 100000  &    4   &   4 & 200 &  8& 36 \\
   10   &  50100  &    4   &   5 & 200 & 10& 10 \\ 
   12   &  33000  &    7   &   6 & 200 & 12& 10 \\
   14   &  25911  &    4   &   7 & 195 & 14& 10 \\
   16   &  18207  &    7   &   8 & 155 & 16& 13  \\
   20   &  10950  &    4   &   9 & 120 & 18&\CC  8 \\
   24   &   9094  &    7   &  10 & 105 & 20&\CC  3 \\
   28   &   7504  &    7   &  11 & 105 & 22&\CC  7\\
   32   &   4811  &    7   &  12 & 100 &  24&\CC  3\\
   40   &   3105  &    7   &  13 & \CC 56 &28&\CC  6\\
   48   &   2620  &    7   &  14 & \CC 51 & 32&\CC  6\\ 
   56   &   1501  &    7   &  15 & \CC 50 & 36&\CC  5\\
   64   &   1536  &    7   &  16 & \CC 50 &40&\CC  7\\
   80   &    703  &    7   &  17 & \CC 50 & 48 &\CC  5\\
   96   &    506  &   10   &  18 & \CC 44 & 56 &\CC  2\\ 
  112   &    205  &   10   &  19 & \CC 44 &    &      \\
  128   &    203  &   10   &     &        &    &      \\
\hline
 \end{tabular}
\parbox[t]{.85\textwidth}
 {
 \caption[Simulation parameters and statistics]
 {\label{Statstandard}
   Left part of table: Statistics of the runs with standard action at
   $\beta=0.2216545$.  stat denotes the number of measurements. Two
   measurements were always separated by $k$ updates with the wall
   cluster algorithm.  Middle: Statistics of the multispin-coding
   runs. Per measurement the demons are once refreshed canonically,
   and there are 8 microcanonical updates of the spin-system.
   Last two columns: Number of
   measurements at $(\beta,D)=(0.383245,0.624235)$ in case of the
   spin-1 model. Two measurements were separated by three cluster 
   updates and one Metropolis sweep.
}}
\ET

\subsection{Improved Action}

A first estimate for $\beta_c$ was obtained by locating the 
crossings of $R_i(L)$ with $R_i(2\,L)$. We used lattices of size
$L=4,8,16,32$ and obtained $\beta_c = 0.383245(10)$, with $D$ given by
eq.~(\ref{goodline}), i.e.\ $D_c = 0.624235$. Note that this was still
a preliminary estimate to be refined later.

Monte Carlo simulations were then performed at $\beta = 0.383245$,
fixing $D$ according to eq.~(\ref{goodline}).  We simulated on cubic
lattices with linear extension $L$ ranging from 4 to 56, using the
single cluster algorithm in alternation with a Metropolis procedure to
maintain ergodicity.  Two measurements were separated by three single
cluster updates and one Metropolis sweep.  As for the standard action
we used clusters built with the boundary flip algorithm to obtain
estimates for $Z_a/Z_p$. After each growth of the corresponding
cluster, the work done was also exploited to perform one wall cluster
step as described for the standard action above.  More detailed
information on the lattice sizes and the statistics can be obtained
from table~\ref{Statstandard}.

The runs for the determination of $\beta_c$ took about three months of
CPU on Pentium 166-MMX PCs, while the final production runs consumed a
total of one year CPU on the same type of PC.

\section{Measured Quantities and Basic Data}

The estimates for the basic quantities relevant for the present study
are quoted in a number of tables in the Appendix.  Here is an
overview:

\begin{center}
\begin{tabular}{llc}
quantities & model & table  \\
\hline 
$R_1$, $R_2$, $\chi$, $E_{NN}$ & Standard Action & \ref{ZQXstandard} \\
$R_1$, $R_2$, $\chi    $       & Improved Action        & \ref{ZQX} \\
$E_{NN}$, $E_D$, $E$           & Improved Action    & \ref{Ae1} \\
\hline 
$dR_1/d\beta$, $dR_2/d\beta$ & Standard Action & \ref{deriStan}  \\
$\partial R_1/\partial \beta$, $d R_1/d \beta$, 
$\partial R_2/\partial \beta$, $d R_2/d \beta$ & Improved Action&\ref{deri} \\
\hline 
\end{tabular}
\end{center}
\vskip3mm

In both models we measured $R_1$, $R_2$, which were already defined in
eqs.~(\ref{defR1}) and~(\ref{defR2}).  We also measured the quantities
needed to estimate the derivatives of $R_1$ and $R_2$ with respect to
the couplings.  Note that in case of the spin-1 model,
eq.~(\ref{goodline}) implies that
\be
\frac{d R_i}{d \beta } = 
\frac{\partial R_i}{\partial \beta } 
+ 3 \, \frac{\partial R_i}{\partial D} \, .
\ee
For the sake of table presentation, the derivatives of the $R_i$ were 
multiplied by a factor $f(L)=L^{-1/0.63}$, which compensates for
the leading divergent behaviour with increasing lattice size.  The
susceptibilities are defined through
\be 
\chi = L^3 \, \langle m^2 \rangle \, , 
\ee 
with $m$ being the normalized magnetization, cf.~eq.~(\ref{defmag}).
In the tables, the susceptibilities are multiplied by a factor $1/L^2$.
For the standard action we measured the energy per link, 
\be\label{ENN} 
E_{NN} = \frac{1}{3 L^3} \sum_{<i,j>} s_i s_j \, . 
\ee 
The same quantity was measured also for the spin-1 model. 
Here in addition we determined  estimates for 
\be
E_D = \frac{1}{L^3} \sum_i s_i^2 \, , 
\ee
and 
\be\label{EEE}
E = - 3 (E_{NN} - E_D ) \, . 
\ee
The definition of the energy $E$ is such that it is proportional 
to $\frac{d}{d\beta} \ln Z$, where the relation of 
$\beta$ and $D$ is given by eq.~(\ref{goodline}). 

Standard reweighting techniques allow us to access the same set of
observables in some neighbourhood of the simulation point.

\section{Fitting the Data}

Our aim is to test the improvement which can be reached in the
estimates for the phenomenological couplings $R_i$ and the critical
exponents using the improved spin-1 action instead of the standard
action. We shall first describe our general approach to the data
analysis.

\begin{figure}
\begin{center}
\includegraphics[width=12cm]{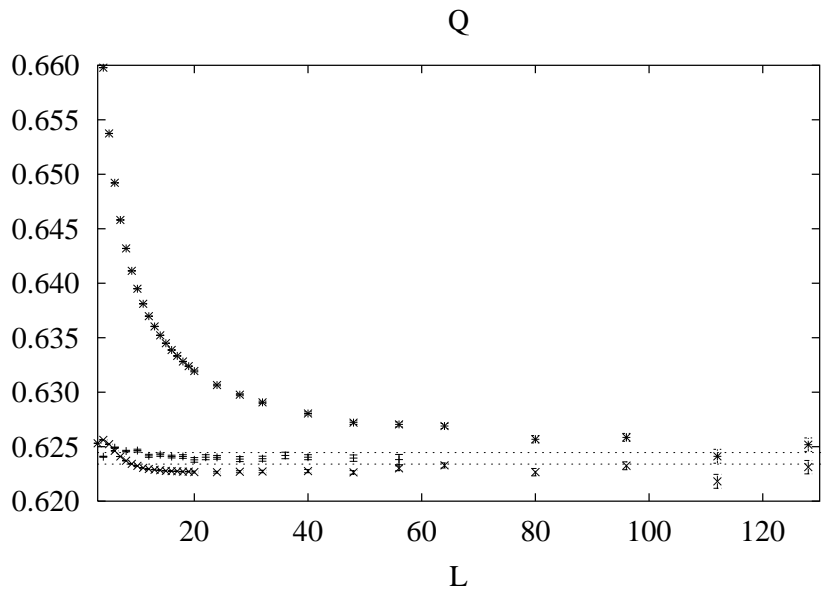}
\PB
 {
 \caption[IA \& SA, plot of $R_2=Q$]
 {\label{figdata2} \small 
   $R_2=Q$ as function of lattice size for both models.  The upper
   data (stars) are from the standard action.  The bars (flat
   data) belong to the improved action.  Crosses  are used to show standard
   action results corrected with the leading correction to scaling
   contribution.  The final estimate obtained from the fit analysis is
   plotted by dotted lines.
 }}
\end{center}
\end{figure}

\begin{figure}
\begin{center}
\includegraphics[width=12cm]{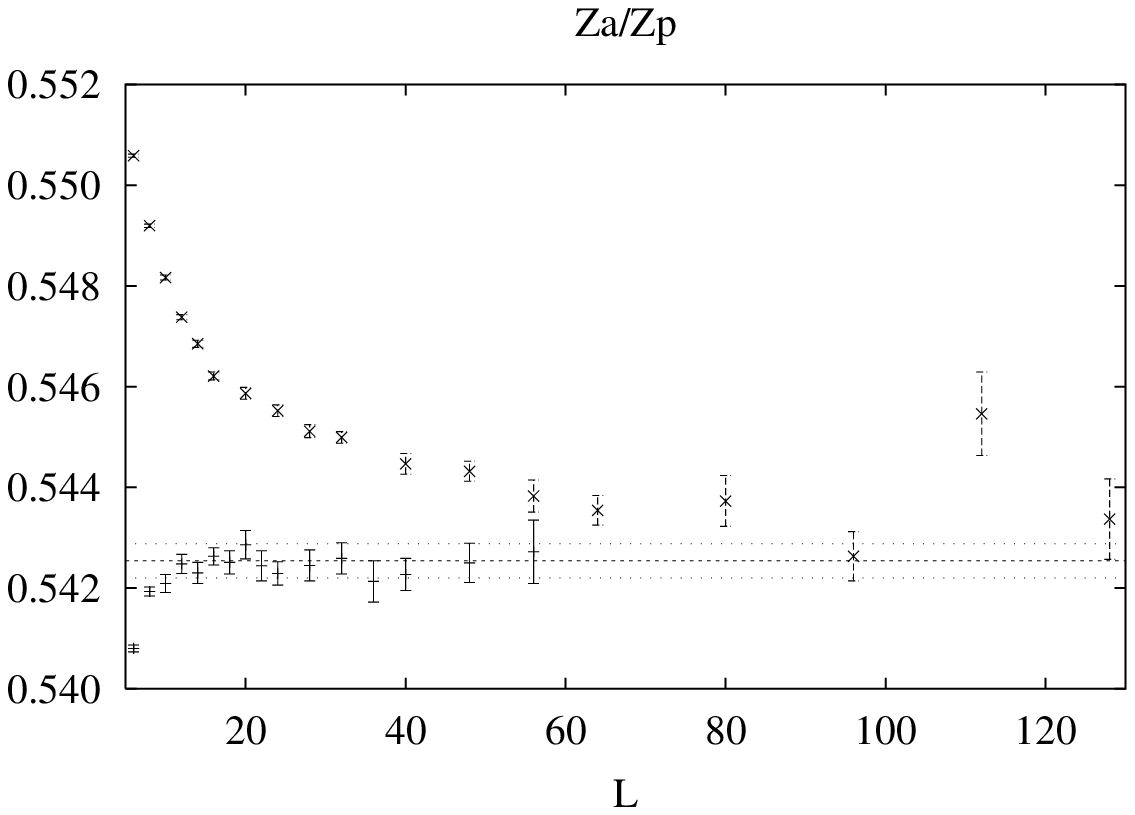}
\PB
 {
 \caption[IA \& SA, plot of $R_1=Z_a/Z_p$]
 {\label{figdata1} \small 
   $R_1=Z_a/Z_p$ as function of lattice size for both models.
   Standard action: crosses, improved action: bars.  The final
   estimate obtained from our fit analysis is also plotted with an 
   error interval.
 }}
\end{center}
\end{figure}

We shall present results obtained from fitting our data for the
standard action and the improved spin-1 action to various finite size
scaling laws.  It will turn out that the estimates obtained from the
standard action are always compatible with those extracted from the
improved action.

In order to get a first impression of the degree of improvement that
can be obtained, the reader is invited to look directly at the tables
of the Appendix (table~\ref{ZQXstandard} to table~\ref{deri}).  One
can easily verify the impressively different behaviour of the standard
and improved data, respectively.

\begin{figure}
\begin{center}
\includegraphics[width=12cm]{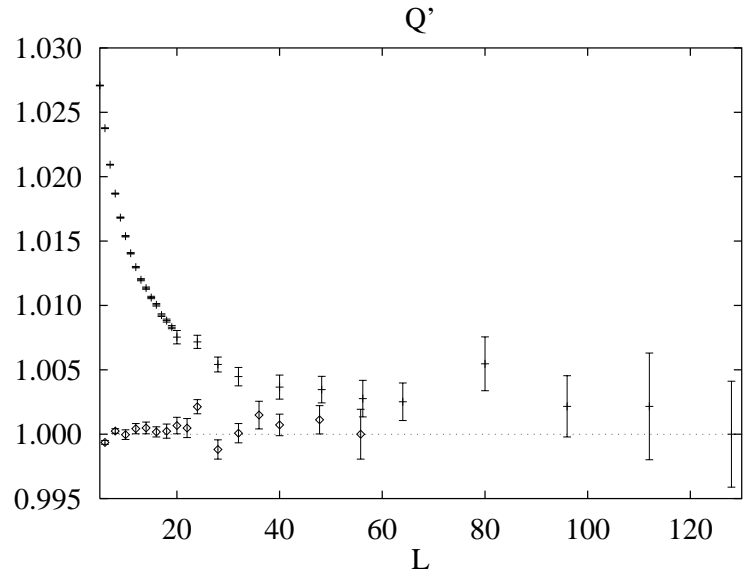}
\includegraphics[width=12cm]{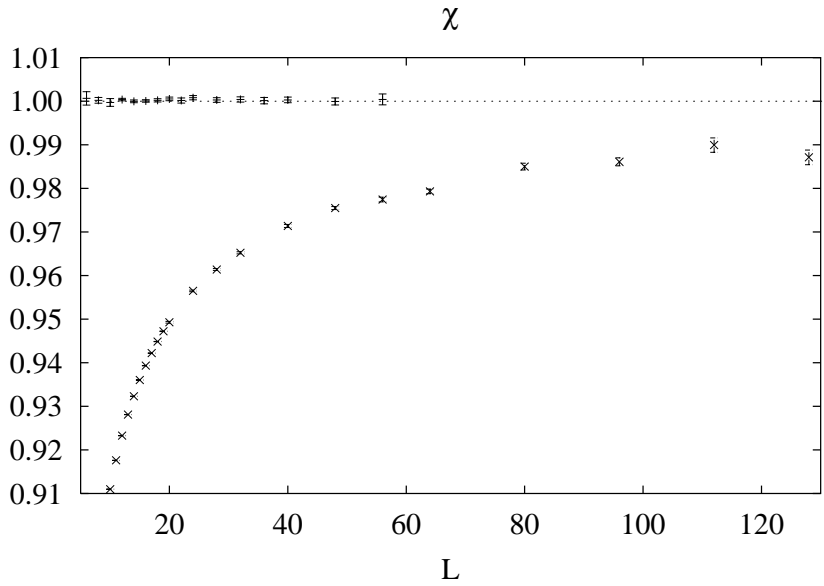}
\PB
{
\caption[IA \& SA, plot of $\beta$ derivatives of $Q$ and 
          plot of $\chi$]
 {\label{figdata3} \small 
   Rescaled $\beta$ derivatives of $Q$ and rescaled $\chi$ at fixed $Q$ as
   function of lattice size for both models (see text).
}}
\end{center}
\end{figure}

In figure~\ref{figdata2}, the Binder cumulants $Q$ are plotted for
the two actions, while in figure~\ref{figdata1} the critical ratios
$Z_a/Z_p$ are given (tables~\ref{ZQXstandard} and \ref{ZQX}).

In both figures, the final values obtained from our fit analysis
are also given, together with error lines indicating
the range of the estimated statistical plus the systematic error.
Clearly the improved samples are much more stable and reach their
asymptotic regime for very small lattice sizes, while the standard
samples hardly get there on lattice sizes of order $10^2$.  In
figure~\ref{figdata2} we have in addition plotted the standard action
result with the contribution of the leading order correction to
scaling subtracted, namely $Q - 0.105 \, L^{-0.81}$, see
table~\ref{BZ1}.

The improved action data outperform the corrected standard action
data.  This is due to the fact that our improvement procedure not only
eliminates correction terms of the form $L^{-\omega}$, but also higher
order corrections of the from $L^{- n \, \omega}$, $n$ integer, which
are generated by the same scaling field.  In particular corrections of
the type $L^{-2 \, \omega}$ should be present in the observables of
the standard Ising model.  Interestingly, the corresponding
corrections have not been taken into account e.g.\ in the analysis by
Bl\"ote et al.~\cite{bloete}, in spite of the fact that $2 \, \omega$
is smaller than other exponents taken into account in the fit
ans\"atze.

Because of their importance in extracting the critical exponents $\nu$
and $\eta$, in figure~\ref{figdata3} we give also the derivatives of
$Q$ and the susceptibilities $\chi$.  In order to check for a residual
$L$-dependence and to be able to compare the samples of the two models,
the data have been rescaled.

The $Q$-derivatives have been multiplied by a factor $L^{-1/0.63}$
(see tables~\ref{deriStan} and \ref{deri}) and normalized to their
value at $L=56$ and $L=128$, respectively, for the improved and
standard action.

Anticipating some results which will be given below, the
$\chi$'s have been transformed by $\chi \rightarrow \left(\chi-
c\right) \, L^{\eta-2}  d^{-1}$, where $ c$, $\eta$, and
$d$ are fit parameters.  The values of the various amplitudes
were taken from tables~\ref{Bchi1} and~\ref{ccc2}.
 
The authors of ref.~\cite{ball1,ball2} claim that the apparent
improvement in the scaling behaviour that we demonstrated above
does not lead to reduced errors in final results for critical exponents. 
Taking the data of the improved model by itself they are in fact
right. Since the coupling parameters of the model are
determined numerically one has to expect some small residual $L^{-\omega}$
corrections, which lead to systematic errors when not taken into 
account in the fit ans\"atze. 

However, we shall demonstrate that it is well possible to estimate the
effects of residual corrections to scaling in a systematic way. We
shall exploit the fact that ratios of correction to scaling amplitudes
are universal.  Given the parametrization
\be
R_i(L)= R^*_i + r_{i} \, L^{-\omega} + \dots  \; ,
\ee
and
\be 
\frac{d R_i}{d\beta} =c_i L^{1/\nu}  (1+ b_i \, L^{-\omega} + \dots  \; ) \, ,
\ee
the ratios $r_{i}/r_{j}$ and $b_{i}/r_{j}$ are universal, i.e., do not
depend on the details of the models chosen \cite{ampli}.
In particular, they are the same for the standard and improved action. 
These ratios
can be obtained from the analysis of the standard Ising model data and
then used in order to estimate the residual correction to scaling
amplitudes in the improved action results.

\subsection{$R_2$ at fixed $R_1$}

Analysing the standard model data, it turns out that the Binder
cumulant $Q$ evaluated at a fixed value of $Z_a/Z_p$ is the optimal
(at least concerning the observables that we measured) quantity to
detect corrections to scaling.\footnote{Doing finite size scaling with
quantities taken at $\beta$-values where a phenomenological coupling
is kept to a fixed value is inspired by ref.~\cite{ballold}.}
Optimal means here that the relative statistical error of the leading
correction to scaling amplitude is the smallest.  We computed $Q$ at
$Z_a/Z_p=0.5425$. This means that first the $\beta(L)$ is computed
where $Z_a/Z_p$ takes the value $0.5425$. Then $Q$ is computed for
these $\beta(L)$. In the following we use $\bar Q$ to denote this
quantity.  In principle $Z_a/Z_p$ could be fixed to any value, however
for practical reasons it is preferable to take a good approximation of
$(Z_a/Z_p)^*$.  To leading order $\bar Q$ should behave as
 \be
 \bar Q = Q^* + r L^{-\omega} + \dots  \; .
 \ee
First we fitted the data obtained with the improved action. We used 
as input $\omega=0.81$ from ref.~\cite{guida}.
\BT{r|c|c|c}
\mc{4}{c}{\bf Improved Action} \\[2mm]
$ L_{\rm min}$ & $ Q^*$ & 
 $r$ &   $\chi^2$/dof \\
\hline
 6& 0.62369(13) &  0.0039(10)  &  1.16 \\
 8& 0.62369(12) &  0.0040(11)  &  1.24 \\
 10&0.62362(14) &  0.0047(13)  &  1.29 \\
 12&0.62371(16) &  0.0036(16)  &  0.50 \\
 \hline
\end{tabular}
\PB
{
\caption[IA, fit $\bar Q$ with correction to scaling term]
{\label{detectI} \small
 Fit of $\bar Q$ with correction to scaling term
 }}
\ET
The results are given in table~\ref{detectI}.  We see that there is
still a small amplitude for corrections to scaling present in the
data. The value of $D$ where the leading order corrections to scaling
exactly vanish should be slightly larger than the one used in this
study.

In order to quantify the improvement that is achieved we have to
compare with the data from the standard Ising model. To obtain a
consistent result of $Q^*$ from small lattices we had to include
a subleading correction to scaling term $ r' \, L^{-2 \omega}$ in the
ansatz. The results are given in table~\ref{detectS}.

\BT{r|c|c|c|c}
\mc{5}{c}{\bf Standard Action} \\[2mm]
$ L_{\rm min}$ & $Q^*$ &
$r$ &$r'$ & $\chi^2$/dof \\
\hline
 6& 0.62294(8)\mx & 0.1191(11) & 0.0574(33) &1.73 \\
 8& 0.62326(12)& 0.1131(19) & 0.0825(67) &0.76  \\
10& 0.62343(14)& 0.1092(27) & 0.102(12)\mx  &0.60 \\
12& 0.62329(18)& 0.1128(43) & 0.081(24)\mx  &0.56  \\
\hline
\end{tabular}
\PB
{
\caption[SA, fit $\bar Q$ with correction to scaling terms]
{\label{detectS} \small
 Fit of $\bar Q$ with correction to scaling terms
}}
\ET

Starting from $L_{\rm min}=8$ the fits have a small $\chi^2$/dof.  The
result for $Q^*$ obtained from $L_{\rm min}=10$ is consistent
with the results from the improved action. There is a clear signal for
the leading order corrections to scaling. The corresponding amplitude
is stable when $L_{\rm min}$ is varied.  Also subleading corrections
are well visible.

Concerning the comparison of the two models we conclude that leading
corrections to scaling in the improved model are reduced by a factor
of about $0.11/0.004=28$ compared with the standard Ising model.  In
order to compute systematic errors due to neglecting $L^{-\omega}$
corrections in the analysis of the data obtained from the improved
model we assume (taking into account the errors in the amplitudes)
that leading corrections to scaling are reduced at least by a factor
of 22 compared with the standard Ising model. 
The reduction from 28 to 22 takes into account that we know the ratio 
of the $r$'s only up to some statistical error.

Using the universality of ratios of corrections to scaling amplitudes 
this reduction has to be the same for all quantities. Hence we can 
take the correction 
amplitudes obtained form the standard action and divide it 
by 22 to obtain a bound on the leading order 
corrections that are to be expected 
in the case of the improved model.

\subsection{Fitting $R_1$ and $R_2$}

We first fitted the $R_i$ in order to obtain estimates for the
phenomenological couplings $R_i^*$, and, in addition, for the critical
coupling $\beta_c$.  Here and in the following we use as an estimate of
the leading correction to scaling exponent $\omega=0.81(2)$ from
ref.~\cite{guida}.

\subsubsection{Standard Action, Fit $R_i$}

\BT{c|l|l|l|c}
\mc{5}{c}{\bf Standard Action}\\[2mm]
  $ L_{\rm min}$ & \phantom{xx} $R^*_1$ 
& \phantom{xx} $\beta_c$ & \phantom{xx} $r_1$ 
 & $\chi^2$/dof  \\
 \hline
 10&0.54275(14)\{49\}  &0.22165446(16)\{21\} & 0.0347(10)  &  1.81\\
 14&0.54304(23)\{44\}  &0.22165431(19)\{15\}*& 0.0318(22)  & 1.88\\
 16&0.54334(26)\{27\}* &0.22165417(18)\{11\} & 0.0281(26)  &1.69 \\
 20&0.54269(40)\{24\}  &0.22165443(21)\{7\}  & 0.0369(50)  &1.43 \\
\hline
\mc{5}{c}{ } \\
  $ L_{\rm min}$ & \phantom{xxx} $R^*_2$  & \phantom{xx} $\beta_c$ 
  & \phantom{xxx} $r_2$  &   $\chi^2$/dof \\
 \hline
 12 &0.62201(6)\{79\} &0.22165363(16)\{64\} &0.11166(44)  &  1.30 \\
 16 &0.62244(16)\{40\}  &0.22165405(23)\{25\}* &0.1077(15)   &  1.01 \\
 20 &0.62292(31)\{21\}*  &0.22165440(28)\{9\} &0.1017(38)   &  1.09 \\
 28 &0.62321(61)\{12\}  &0.22165457(40)\{4\} &0.0973(87)   &  1.33 \\
 \hline
 \end{tabular}
\PB
 {
 \caption[SA, fit $R_i$ separately, with $\omega$ fixed]
 {\label{BZ1} \small 
Fitting separately the $R_i$ with eq.~(\ref{fitR3}), 
fixing $\omega=0.81$.
}}
\ET

In the case of the standard action we fitted our data with the ansatz
\be
\label{fitR3}
R_i(L,\beta_{\rm MC})= R_i^* + \frac{d R_i}{d\beta}(L,\beta_{MC})
\, \Delta \beta \, + r_i \, L^{-\omega} \, . 
\ee
In addition to a correction to scaling term of the form $r_i \,
L^{-\omega} $ we included a term which (to first order) corrects for
deviations from being at criticality.  $\Delta \beta$ is the
difference between the critical $\beta_c$ and $\beta=0.2216545$.
Fitting $R_1$ to this law, we first fixed $\omega= 0.81$.  The fits
were done on a sequence of data sets obtained by discarding data with
$L < L_{\rm min}$.  The results for the fit parameters as function of
the smallest lattice size included are given in table~\ref{BZ1}.  The
procedure used to compute the estimates of the systematic errors
(curly brackets) is discussed at the end of this section.  In the
table we mark with an asterix the value of $L_{\rm min}$ where the
systematic error estimate becomes equal or smaller than the
statistical estimate.

\begin{figure}
\begin{center}
\includegraphics[width=12cm]{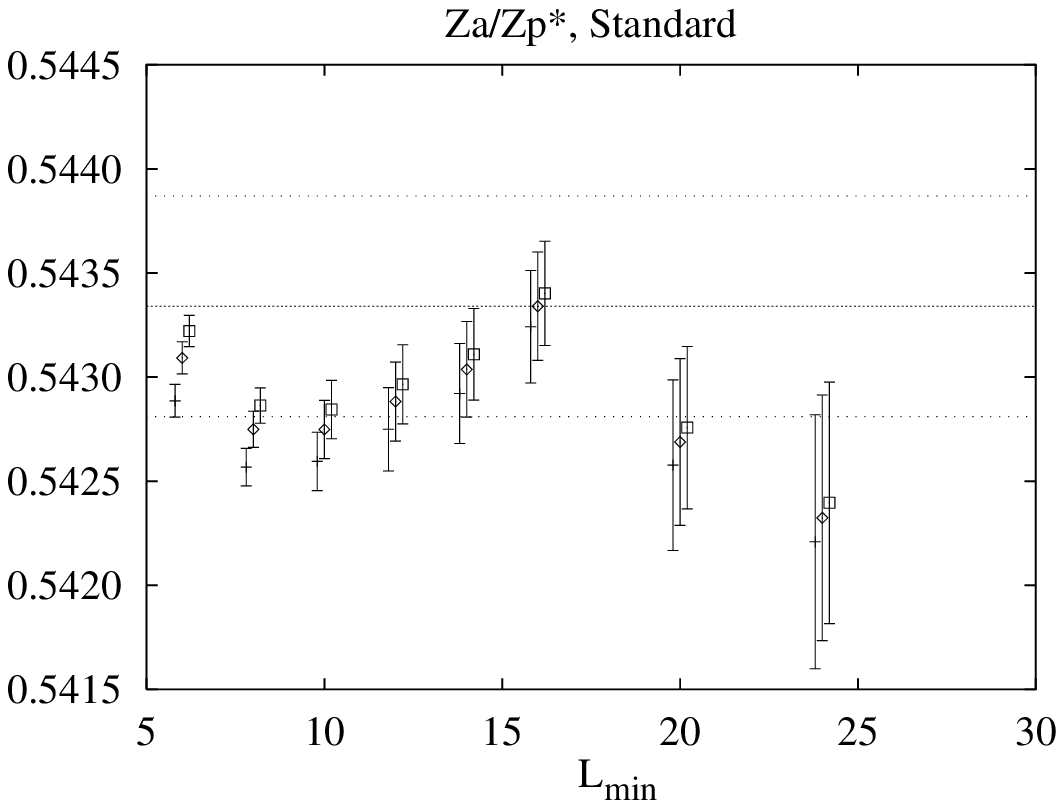}
\includegraphics[width=12cm]{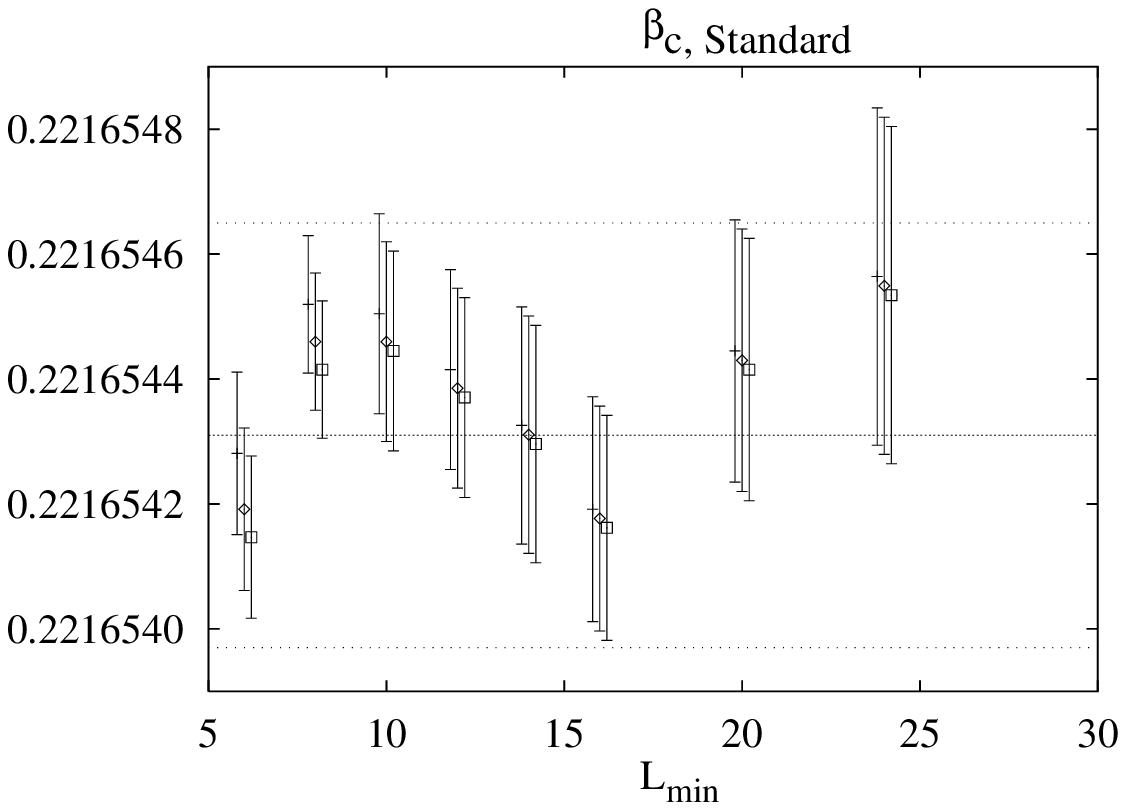}
\parbox[t]{.85\textwidth}
 {
 \caption[SA, $R^*_1$ and $\beta_c$ from fit with fixed $\omega$]
 {\label{figZSt} \small
   $R_1^*$ and $\beta_c$, standard action, 
   from fits to eq.~(\ref{fitR3}), with fixed
   exponents $\omega=0.78, 0.81, 0.83$.  
 }}
\end{center}
\end{figure}

The fits are reasonably stable.  Figure~\ref{figZSt} shows the fit
results for $R_1^*$ and $\beta_c$ as function of $L_{\rm min}$.  In
addition, the figure gives an impression on the dependence of the
estimates on the choice of the fixed $\omega$, by varying its value
through 0.78 to 0.83.  The conclusion is that the dependence on the
choice of $\omega$ is negligible compared to the statistical errors
and the systematic errors quoted in table~\ref{BZ1}.
Note that here and in the following all error ranges 
for final fit results in the plots (usually indicated by a set of 
horizontal lines) are obtained by {\em adding}
the statistical and systematic errors.  

\begin{figure}
\begin{center}
\includegraphics[width=12cm]{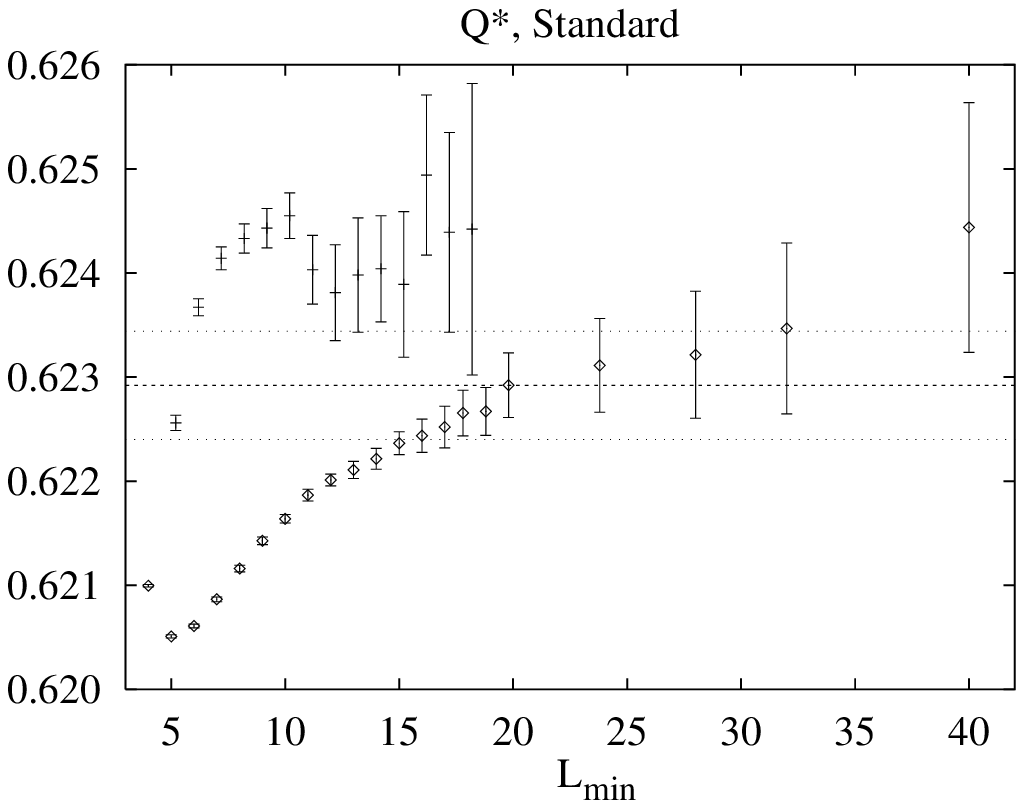}
\includegraphics[width=12cm]{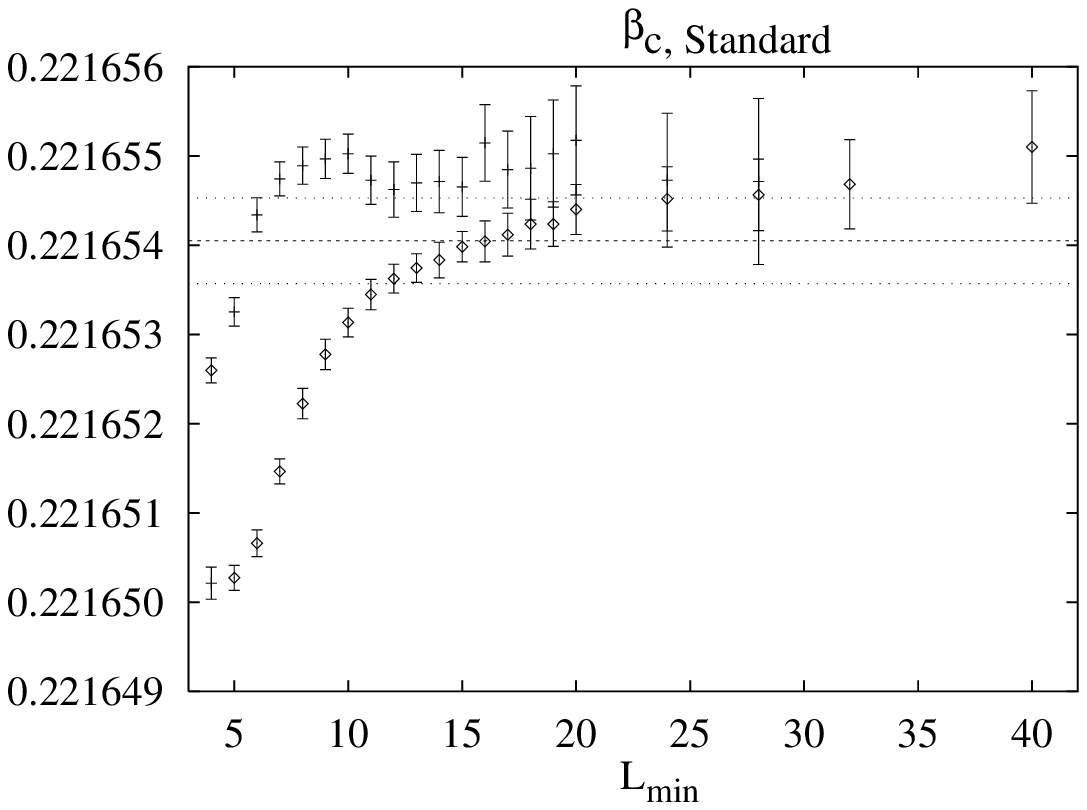}
\PB
 {
 \caption[SA, $R_2^*$ and $\beta_c$, with $\omega$ fixed and $\omega$ free]
 {\label{figQSt}
\small
$R_2^*$ and $\beta_c$,
standard action, from fits  with eq.~(\ref{fitR3}).
Diamonds: $\omega=0.81$ fixed. Bars: $\omega$ free fit parameter.
}}
\end{center}
\end{figure}

Let us now turn to the cumulant $R_2$. The results are given in
table~\ref{BZ1} and in figure~\ref{figQSt}. In the figure also the
results of table~\ref{BQ1}, where we have let $\omega$ a free
parameter, are shown.  The outcome of $\omega$ is significantly larger
than 0.81, which we presently accept as a reliable estimate.  We
interpret this as a sign that this fit parameter tries to compensate
for the lack of even higher correction exponents.

\BT{r|l|l|l|c|c}
\mc{6}{c}{\bf Standard Action} \\[2mm]
  $ L_{\rm min}$ & \phantom{xxx} $R^*_2$  & \phantom{xxx} $r_2$ 
  & \phantom{xxx} $\omega$ &$\beta_c$ &  $\chi^2$/dof \\
 \hline
  6 &0.623670(82)& 0.13732(51) &0.9384(37)& 0.22165434(19) & 2.60  \\
  8 &0.62433(14) & 0.1449(18)  &0.9802(93)& 0.22165489(21) & 0.89 \\
 10 &0.62455(22) & 0.1490(40)  &0.998(18)& 0.22165503(22) & 0.94 \\
 16 &0.62494(77) & 0.174(20)   &1.070(77)& 0.22165515(43) & 0.92 \\
 20 &0.6252(22)  & 0.22(11)    &1.16(25)& 0.22165517(61) & 1.12 \\
 \hline
 \end{tabular}
\PB
{
\caption[SA, fit $R_2$, with $\omega$ free]
{\label{BQ1} \small 
Fit of $R_2$ with eq.~(\ref{fitR3}).
}}
\ET

\BT{l|l|l|l|c}
\mc{5}{c}{\bf Standard Action}\\[2mm]
\phantom{x} $x$ &
\phantom{xx} $R^*_1$&  \phantom{xx}$R^*_2$
& \phantom{xx}$\beta_c$ & $\omega$ \\
\hline
1.62    &0.54120(19)&0.62295(25) &0.22165516(11) & 0.670(37) \\
2.0     &0.54256(15)&0.62261(25) &0.22165443(11) & 0.772(31) \\
2.4     &0.54305(13)&0.62325(24) &0.22165437(11) & 0.865(27) \\
2.30(41)&0.54285(31)&0.62261(29) &0.22165428(14) & 0.800(58) \\
\hline
final   & 0.54334(26)\{27\}  &0.62292(31)\{21\} & 0.22165431(19)\{15\} &   \\
\hline 
 \end{tabular}
 \begin{tabular}{l|c|l|l|c} 
\mc{4}{c}{ } \\
\phantom{x} $x$ & $r_1$ & \phantom{xx} $r_2$ & \phantom{xx} $r'_2$
&  $\chi^2$/dof \\ 
 \hline
1.62    & 0.0324(20) & 0.058(11) & 0.168(19)  & 4.40 \\
2.0     & 0.0330(19) & 0.0907(87) & 0.152(25)  & 1.60 \\
2.4     & 0.0373(17) & 0.1151(72) & 0.136(36) & 1.31  \\
2.30(41)& 0.0334(28) & 0.100(32)  & 0.184(25) & 1.63  \\
 \hline
 \end{tabular}
\parbox[t]{.85\textwidth}
 {
 \caption[SA, fit $R_i$ simultaneously, with $\omega$ free and extra 
         $x$ for $R_2$]
 {\label{BZQ1} \small 
Fitting together $R_1$ and $R_2$ for $L_{\rm min}=8$.
For $R_1$ we used eq.~(\ref{fitR3}), while 
for $R_2$ also an effective next-to-leading correction $r'_2\,L^{-x}$
was added.
}}
\ET

We also fitted the two $R_i$ together.  We did this by allowing
$\omega$ to vary freely.  For $R_1$, the exponent $\omega$ is the only
correction to scaling, whereas for $R_2$ we included additionally a
correction term with exponent $x$.  We checked the two possibilities
to either fix the exponent $x$ to some value or subject it to fitting
in order to assume some effective value.  Indeed, beyond the leading
correction to scaling exponent $\omega$, there are several exponents
that could enter the game, like for instance 
$2\omega$, $x\approx 2$, or $1/\nu+\omega$.
In table~\ref{BZQ1} we present the fit results.

Notice that the amplitude of the leading correction to scaling is much
smaller for $R_1$, as expected, and for $R_2$ also the amplitude of
the next-to-leading correction to scaling $r_2'$ is significantly
different from zero.  The effective next-to-leading exponent is of
order two, and actually compatible with that appearing in the improved
case. This could well be a coincidence.

Finally, the errors of this fit are smaller than the (safer) errors
obtained from the fits with fixed $\omega$.  However, as a consequence
of the limited number of lattice sizes available, the fit with many
parameters can not be checked for stability with respect to varying
$L_{\rm min}$.

In order to get a handle on the systematic errors due to the
neglection of higher order corrections, we used the result from the
previous simultaneous fit of $R_1$ and $R_2$ to estimate the effective
next-to-leading exponents.  We assume this exponent $x$ to be of order
2.  Notice that from table~\ref{BZQ1} it turns out that the value of
$r_2'$ does not depend strongly on $x$. The same holds for $r_1'$,
which is estimated to be of order 0.02 by a 4-parameter fit with
$\omega=0.81$, $x=2.3$, on $Z_a/Z_p$.

Then we fitted separately to eq.~(\ref{fitR3}) the quantities defined by 
 \be
 \label{NTL}
 \tilde R_i(L) =  R_i(L) - r'_i L^{-x}  \; ,
 \ee
where $R_i(L)$ are the original (Monte Carlo) data, and $r'_i$ and $x$
have fixed values determined by the fits discussed above, namely 
$ r_1'=0.02$, $r_2'=0.184$, and $x=2.3$.

Roughly speaking, the $\tilde R_i(L)$ are the original data after subtraction 
of an estimate of the subleading correction to scaling contamination.

The absolute values of the differences between the $R_i^*$ and
$\beta_c$ obtained in this way and those obtained from fitting
$R_i(L)$ with the same equation are the estimates of the systematic
errors given (inside curly brackets) in table~\ref{BZ1}.

In summary, from  table~\ref{BZ1} we obtain the following final estimates
(labelled with a * in the table):

\begin{center}
\begin{tabular}{rrl}
                       &  $R_1^*$    &=~0.54334(26)\{27\}   \\[1mm]
{\bf standard action:} &  $R_2^*$    &=~0.62292(31)\{21\}  \\[1mm]
                       &  $\beta_c$  &=~0.22165431(19)\{15\} from $R_1$ \\[1mm]
                       &  $\beta_c$  &=~0.22165405(23)\{25\} from $R_2$ 
\end{tabular}
\end{center}

The final estimates together with their error-lines are also 
given in figures~\ref{figZSt} and \ref{figQSt}. 

\subsubsection{Improved Action, Fit $R_i$}

\BT{r|l|l|c}
\mc{4}{c}{\bf Improved Action} \\[2mm]
$ L_{\rm min}$ & \phantom{xx} $R_1^*$  &
\phantom{xxx} $\beta_c$ &   $\chi^2$/dof \\
 \hline
  8& 0.54213(8) [24]\{66\}& 0.3832470(8)[8]\{36\} &   1.78 \\
 10& 0.54240(10)[19]\{32\}& 0.3832453(9)[5]\{15\} &   0.79 \\
 12& 0.54251(11)[17]\{21\}& 0.3832447(9)[4] \{9\}* &   0.49 \\
 16& 0.54260(15)[16]\{15\}*& 0.3832442(11)[4]\{5\} &   0.46 \\
 20& 0.54252(25)[13]\{10\}& 0.3832446(14)[2]\{3\} &   0.55 \\
\hline 
\mc{4}{c}{} \\
$ L_{\rm min}$ & \phantom{xx} $R_2^*$  &
\phantom{xxx} $\beta_c$ &   $\chi^2$/dof \\
\hline 
  8& 0.62447(6)[76]\{30\}& 0.3832499(10)[46]\{29\}&  2.22 \\
 10& 0.62429(7)[62]\{15\}& 0.3832479(12)[31]\{12\}&  1.39 \\
 12& 0.62414(8)[57]\{11\}& 0.3832465(11)[26]\{9\}&  0.42 \\
 16& 0.62405(12)[50]\{7\}& 0.3832457(14)[20]\{5\}&  0.31 \\
 20& 0.62393(18)[43]\{4\}& 0.3832447(18)[17]\{3\}*&  0.28  \\
 \hline
 \end{tabular}
\PB 
{
\caption[IA, fit $R_i$ separately, no correction to scaling]
{\label{RRR2} \small
Fitting separately the $R_i$ with eq.~(\ref{fitR}). The numbers
in square and curly brackets are estimates of the systematic errors
(see text).
}}
\ET

\BT{r|l|l|l|c}
\mc{5}{c}{\bf Improved Action} \\[2mm]
$ L_{\rm min}$ & \phantom{xx} $R_1^*$  & \phantom{xx} $R_2^*$  &
\phantom{xxx} $\beta_c$ &   $\chi^2$/dof \\
\hline
8&0.54206(7)[11]\{65\}&0.62441(5)[64]\{32\} & 0.3832481(6)[13]\{34\}&  2.19 \\
10&0.54231(9)[7]\{32\}&0.62421(6)[51]\{16\} & 0.3832463(7)[8]\{14\}&  1.21 \\
12&0.54245(11)[7]\{22\}&0.62408(7)[46]\{11\} & 0.3832453(8)[6]\{9\} &  0.51 \\
16&0.54254(14)[6]\{14\}*&0.62399(8)[40]\{7\}  & 0.3832448(9)[6]\{5\}*&  0.42 \\
20&0.54252(22)[4]\{9\}& 0.62393(13)[35]\{5\} & 0.3832446(11)[4]\{3\}&  0.41 \\
 \hline
 \end{tabular}
\PB
{
\caption[IA, fit $R_i$ simultaneously, no correction to scaling]
{\label{RRRb} \small 
  Fitting simultaneously the $R_i$ with eq.~(\ref{fitR}).
}}
\ET

We fitted the data to  
\be  
\label{fitR}
R_i(L,\beta_{\rm MC})= R_i^* + \frac{d R_i}{d\beta}(L,\beta_{\rm MC})
\, \Delta \beta \, . 
\ee
Again we included a term which (to first order) corrects for deviations from
being at criticality. $\beta_{\rm MC}$ is our simulation coupling
0.383245, and the $d R_i/d\beta$ are taken from table~\ref{deri}.  The
fit parameters are $R_i^*$ and $\beta_c$, entering through $\Delta
\beta= \beta_{\rm MC}-\beta_c$.  
We first fitted separately $R_1$ and $R_2$ in order to compare their
scaling behaviour. The results are reported in table~\ref{RRR2}.  
For both quantities the fit parameter estimates are quite stable.

Here and in the following, for what concerns estimates obtained from
the improved action, in addition to the usual error bars (quoted in
usual brackets), we quote for each fit parameter two systematic
errors.  The systematic errors are meant as an estimate of the
uncertainty due to corrections to scaling terms.  The first one,
square brackets, estimates the error made neglecting a leading
correction scaling term.  The second one, curly brackets, as in the
standard action case, estimates the error made neglecting higher
subleading corrections to scaling.  They were obtained in a well
defined way to be described at the end of this section.

We also fitted all the $R_i$-data together with three parameters
($R_1^*$, $R_2^*$, and $\beta_c$).  The results are presented in
table~\ref{RRRb}. 

\begin{figure}
\begin{center}
\includegraphics[width=12cm]{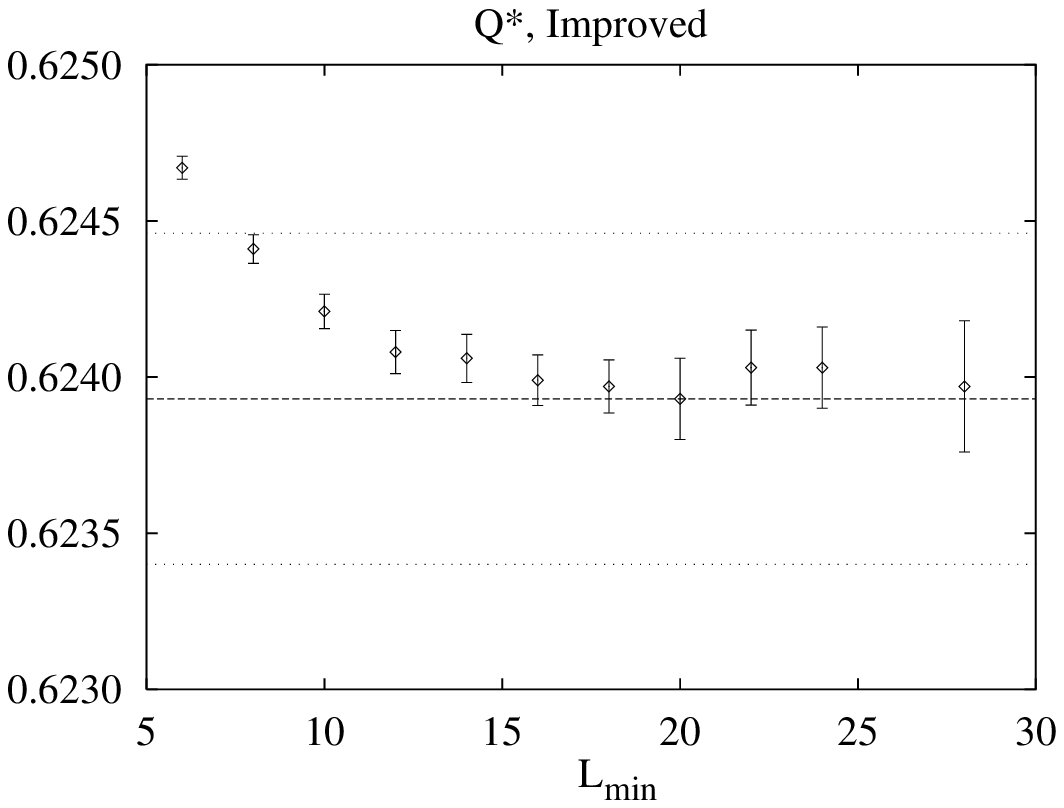}
\includegraphics[width=12cm]{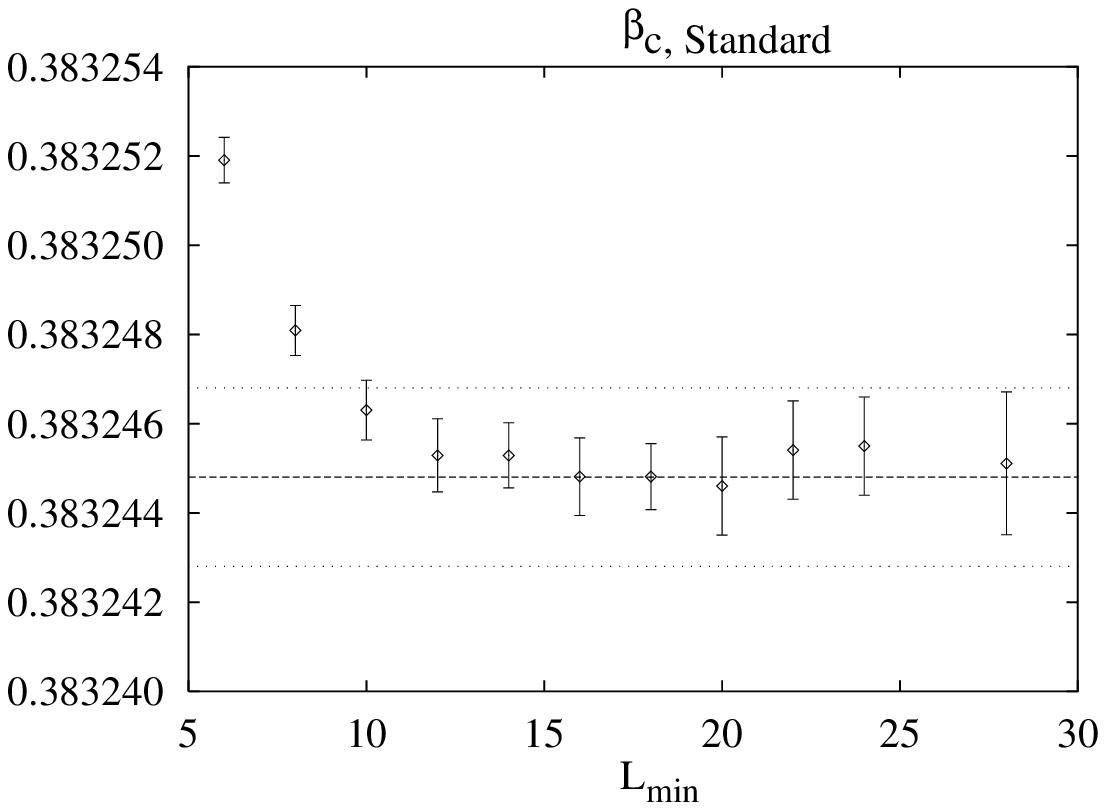}
\PB
 {
 \caption[IA, $R^*_2$ and $\beta_c$ from simultaneous fit of $R_i$]
 {\label{figQ} \small 
 $R_2^*$ and $\beta_c$ as function of smallest lattice size included 
 in a common fit of $R_1$ and $R_2$ data to eq.~(\ref{fitR}).
 }}
\end{center}
\end{figure}

%Note that the estimate for $\beta_c$ is consistent with the
%preliminary estimate from the crossing method, c.f.\ section~3,
%however, with an error reduced by a factor of 10.

Before describing the procedure we used to estimate systematic errors 
for the improved action case, let us report the results from a
simultaneous fit of $R_1$ and $R_2$ with the law
\be
\label{fitR2}
R_i(L,\beta_{\rm MC})= R_i^* + \frac{d R_i}{d\beta}(L,\beta_{\rm MC})
\, \Delta \beta \, + r_i \, L^{-x} \, . 
\ee
As already discussed in the previous section, $x$ represents an effective
exponent.

\BT{l|l|l|l|c}
\mc{5}{c}{\bf Improved Action}\\[2mm]
\phantom{x} $x$& 
\phantom{xx}$R^*_1$ & 
\phantom{xx}$R^*_2$ & \phantom{xx} $\beta_c$ & $\chi^2$/dof \\
\hline
0.81     &0.54423(23)    &  0.62314(13)   &0.383241(1)  &  1.52 \\
2.0\mx   &0.54287(12)    &  0.62394(7)\mx &0.383244(1)  &  1.00 \\
2.5\mx   &0.54268(11)    &  0.62405(7)\mx &0.3832446(8) &  0.96 \\
2.46(56) &0.54269(17)    &  0.62405(13)   &0.3832446(11)&  1.00 \\
\hline
final    &0.54254(14)[6]\{14\}&0.62393(13)[35]\{5\}&0.3832448(9)[6]\{5\}& \\
\hline 
 \end{tabular}
\begin{tabular}{l|c|c}
\mc{3}{c}{ } \\
\phantom{x} $x$& $r_1$ & $r_2$ \\
\hline
0.81     &--0.014(1)\mx  &0.0076(6)     \\
2.0      &--0.07(6)\mxx  &0.036(4)\mx   \\
2.5      &--0.162(13)    &0.080(8)\mx   \\
2.46(56) &--0.15(57)\mx  &0.07(22)\mx   \\
\hline
 \end{tabular}
\PB
{\caption[IA, fit $R_i$ simultaneously, with correction to scaling]
{\label{Aqz1} \small 
Fitting simultaneously the $R_i$ with eq.~(\ref{fitR2}), see text.
}}
\ET

Our fit results are summarized in table~\ref{Aqz1}.  In the first
three lines fits with fixed $x$ are shown, while in the fourth $x$ is
a free parameter. All these fits correspond to $L_{\rm min}=6$.  In
the fifth line our final estimates are given for comparison.

If we force the $x$-exponent to assume the leading correction value
$\omega=0.81$, we observe that $\chi^2/$dof is a little larger than for 
the other values of $x$. The correction amplitudes $r_i$  
are very small. The main problem of this fit is  
that the ratio $r_1/r_2$
is completely inconsistent with that found for the standard Ising action.

Leaving $x$ free, it tends to choose a value around $2.5$. However the
corresponding fit does not improve on the estimates of the fit without
correction to scaling, though consistent with it. Compare also the
fits with $x$ fixed to $2$ and to $2.5$.

Note that the $\chi^2/$dof values do not allow to discriminate
between the different solutions, i.e.\ the different exponents.
Obtaining estimates from a (relatively) large lattice limit of fits without
correction to scaling is, in this case, a safer 
procedure compared to a multi-parameter fit on all lattice sizes, in 
principle legitimate but in practice difficult to control.

Being the $\chi^2/$dof not a good signal of the presence of correction to 
scaling terms, we estimated the systematic error due to neglecting the leading
correction to scaling term as well as that due to subleading corrections.

For the estimates of errors due to leading corrections to scaling we
used the following procedure (systematically used also in the
following sections).  From table~\ref{BZ1} we know with good precision
the leading correction amplitudes $r^{(S)}_i$ for the standard action.
{}From the universality argument discussed in section~5.1, we assume that
the corresponding amplitudes for the improved action are given by
$r^{(S)}_i/22$.  Therefore, analogously to eq.~(\ref{NTL}), we define
the tilde quantities
 \be
 \label{NTL2}
 \tilde R_i(L) =  R_i(L) - \frac{r^{(S)}_i}{22}\, L^{-\omega}  \; .
 \ee
As in the previous section, fitting $\tilde R_i$ and $R_i$ with 
eq.~(\ref{fitR}) and taking the absolute value of the differences of the 
outcome parameters gives the estimates reported in square brackets.

The information from the fits of table~\ref{Aqz1} with the extra exponent $x$ 
can instead be used to estimate in a systematic way the effect of ignoring
the corresponding corrections. 
We define again
\be
\label{corri}
\tilde R_i(L)=  R_i(L) - r_i\, L^{-x} \, , 
\ee
where $x$ and $r_i$ are parameters obtained from the fits reported in
table~\ref{Aqz1}, namely $r_1= 0.07$, $r_2=-0.15$, and $x=2.46$.
Repeating the steps followed above, one obtains the estimates given in
curly brackets.

Using the fit estimates marked with an asterix, i.e., where
statistical and systematic error estimates are of the same order, we
obtain our final estimates:
\begin{center}
\begin{tabular}{rrl}
                       &  $R_1^*$    & = ~0.54254(14)[6]\{14\}   \\[1mm]
{\bf improved action:} &  $R_2^*$    & = ~0.62393(13)[35]\{5\}  \\[1mm]
                       &  $\beta_c$  & = ~0.3832448(9)[6]\{5\} \\[1mm]
\end{tabular}
\end{center}

The $R_2^*$ and $\beta_c$ (as function of $L_{\rm min}$) are shown in 
figure~\ref{figQ}. In the figures, our final estimates are also 
indicated by dashed lines, with the error intervals bounded by dotted lines. 

\subsection{Fitting the Derivatives of the $R_i$}

\subsubsection{Standard Action, Fit $d R_i /d \beta$}

\BT{r|l|l|c}
\mc{4}{c}{\bf Standard Action }\\[2mm]
 $L_{\rm min}$  & \phantom{xx} $a_1$  & \phantom{xx} $\nu$ &  $\chi^2$/dof \\
 \hline
  10&--1.4723(7)&  0.62981(7)[102]   & 5.38 \\
  20&--1.4798(26)& 0.63045(21)[85] & 1.00 \\
  28&--1.4747(36)& 0.63008(27)[69] & 0.88 \\
  40&--1.4757(79)& 0.63014(54)[55]* & 0.96 \\
  48&--1.4699(13)& 0.62977(88)[48] & 1.09 \\
\hline
\mc{4}{c}{}\\[2mm]
  $L_{\rm min}$ & \phantom{xx}  $a_2$  & \phantom{xx} $\nu$ 
 &  $\chi^2$/dof    \\
 \hline
 14& 0.87050 (71)& 0.6331(1)[31]&3.35\\
 24& 0.8592  (29)& 0.6315(4)[18]&0.66\\
 40 & 0.8501(76) & 0.6305(9)[13] & 0.26 \\
 48 & 0.850(12)  & 0.6304(13)[11]*  & 0.31 \\
 56 & 0.848(14)  & 0.6303(15)[11]  & 0.39 \\
 \hline
 \end{tabular}
\PB 
{\caption[SA, fit $d R_i/d \beta$, no correction to scaling]
{
\label{BZder1} \small  
 Fitting $d R_1/d \beta$ (top) 
 and $d R_2/d \beta$ (bottom) with eq.~(\ref{fitQp}). 
 }}
\ET

We first fit the $d R_i/d\beta$ without correction to scaling,

\be
\label{fitQp}
\frac{ \partial R_i}{\partial \beta} = a_i \, L^{1/\nu} \, .
\ee

The corresponding results are summarized in table~\ref{BZder1}. 
As expected, both quantities suffer from strong corrections to scaling.
Let us first estimate the systematic error due to the leading correction.
We followed the procedure described in section~5.2.
Namely, we made fits with the ansatz
\be\label{Rder2}
\frac{\partial R_i}{\partial \beta}
= a_i \, L^{1/\nu} ( 1 + b_i \, L^{-\omega} )  \, ,
\ee
then we defined 
\be\label{Rder22}
\tilde \frac{\partial R_i}{\partial \beta}
= \frac{\partial R_i}{\partial \beta} - a_i \, b_i \, L^{1/\nu-\omega} \, , 
\ee
and finally we fitted the tilde quantities to eq.~(\ref{fitQp}), 
fixing $\omega=0.81$.
The differences in the $\nu$ exponents are given in the square brackes of 
table~\ref{BZder1}.
The leading amplitude corrections $b_i$ can be found in table~\ref{BQder2} and 
table~\ref{BZder2}.

\begin{figure}
\begin{center}
\includegraphics[width=10cm]{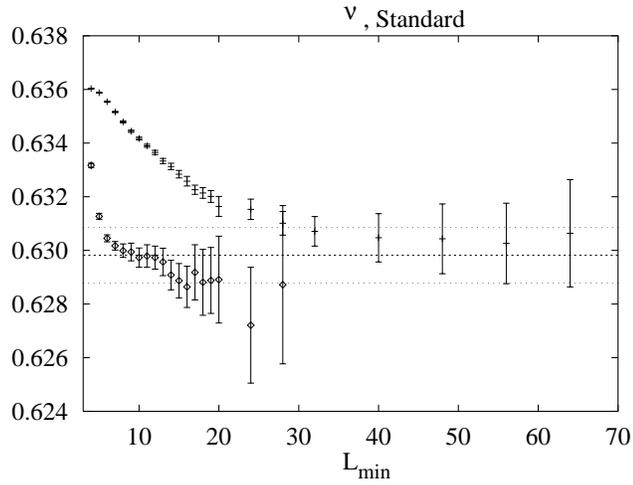}
\PB
 {
 \caption[SA, $\nu$ from $d R_2/d\beta$, without and with
correction to scaling]
 {\label{figQderSt} \small
$\nu$ resulting from fitting $d R_2/d\beta$, standard action, without
correction to scaling (bars) and with $\omega=0.81$ (diamonds).
 }}
\end{center}
\end{figure}

The derivative of $R_2$ suffers from stronger systematic effects than
the $R_1$ derivative.  Therefore we include the $\omega$
correction into the fit ansatz (namely, we fit with eq.~(\ref{Rder2}))
and compute the systematic error made neglecting further subleading
correction to scaling.  The fit results are given in
table~\ref{BQder2} and in figure~\ref{figQderSt}, where also the
$\nu$-exponents obtained from the fit with eq.~(\ref{fitQp}) are
reported for comparison.

In the table we have included as a third error bar (in $\LL \GG$
brackets) estimates of the systematic effect from varying $\omega$
from 0.77 through 0.85. 
This covers a 2-$\sigma$-interval
around the $\omega$ value 0.81(2) that we decide to use.
Again, the systematic error estimates in
curly brackets take into account the omission of next-to-leading
corrections to scaling.  They were computed with the procedure
discussed at the end of section~5.2.1.  We fitted with the ansatz

\be\label{Rder3}
\frac{\partial R_i}{\partial \beta}
= a_i \, L^{1/\nu} ( 1 + b_i \, L^{-\omega} + b'_i \, L^{-x}) \, ,
\ee
and defined the tilde quantities subtracting a contribution 
$a_2 \, b^{'}_2 \, L^{1/\nu-2}$, with an estimate $b_2^{'}=-0.1$.

\BT{r|l|l|l|c}
\mc{5}{c}{\bf Standard Action}\\[2mm]
$L_{\rm min}$ & \phantom{xx} $a_2$  & \phantom{xx} $\nu$& 
\phantom{xx} $b_2$ & $\chi^2$/dof  \\
 \hline
  8& 0.8418(20)  & 0.62999(25)\{93\}$\LL 20\GG$ & 0.1103(59) & 0.48  \\
 10& 0.8396(29)  & 0.62973(36)\{61\}$\LL 17\GG$ & 0.1181(95) & 0.47  \\
 12& 0.8395(36)  & 0.62973(43)\{46\}$\LL 14\GG$* & 0.119(13)  & 0.52  \\
 14& 0.8333(48)  & 0.62908(55)\{37\}$\LL 16\GG$ & 0.147(19)  & 0.35  \\
 \hline
 \end{tabular}
\parbox[t]{.85\textwidth}
 {
 \caption[SA, fit $d R_2 / d\beta$, with $\omega$ fixed]
{\label{BQder2} \small 
 Fitting $d R_2 / d\beta$  with eq.~(\ref{Rder2}) and fixed $\omega=0.81$.
 }}
\end{center}
\end{table}

\BT{c|c|c|c|c|c}
\mc{6}{c}{\bf Standard Action}\\[2mm]
$x$&$a_1$ & $\nu$ & $b_1$ & $b'_1$&$\chi^2$/dof \\
 \hline
1.62&--1.446(12)&0.62873(69)&0.175(44)&--0.65(11) &0.94\\
2.0 &--1.460(10)&0.62943(61)&0.091(32)&--0.83(13) &1.0\\
2.4 &--1.469(9)\mx&0.62993(55)&0.046(26)&--1.27(20) &1.1\\
 \hline
\end{tabular}
\PB 
{\caption[SA, fit $d R_1/d\beta$ with leading and higher correction]
{\label{BZder2} Fitting $d R_1/d\beta$ with eq.~(\ref{Rder3}) and
fixed $\omega=0.81$ for $L_{\rm min}=8$. 
}}
\ET
In  table~\ref{BZder2} we give for comparison the results of fitting $R_1$ to 
eq.~(\ref{Rder3}) for $L_{\rm min}=8$. 
Notice that in this case the data signals the 
presence of a stronger next-to-leading correction to scaling.
The $\nu$ obtained in this way is anyhow consistent with the final estimate 
obtained from the $R_2$ derivative.

We quote as our final estimates 
\begin{center}
\begin{tabular}{rrll}
{\bf standard action:} &$\nu$ &=~  0.63014(54)[55] 
                       &(from table~\ref{BZder1})  \\[1mm]
                       &$\nu$ &=~ 0.62973(43)\{46\}$\LL 14 \GG$ 
                       &(from table~\ref{BQder2})
\end{tabular}
\end{center}

\subsubsection{Improved Action, Fit $\partial R_i/\partial \beta$}

\begin{figure}
\begin{center}
\includegraphics[width=10cm]{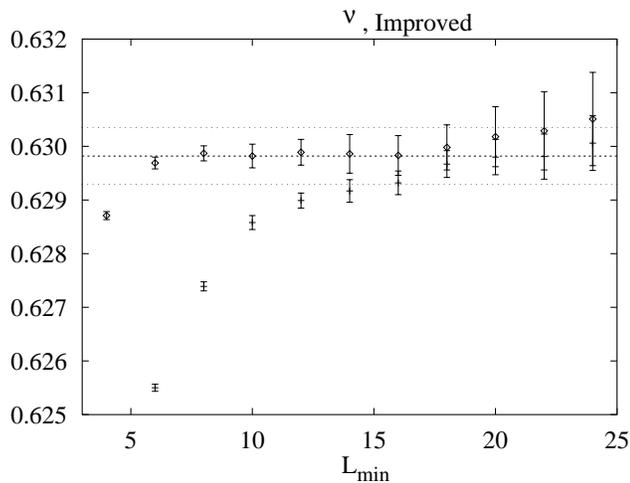}
\parbox[t]{.85\textwidth}
{
\caption[IA, $\nu$ from $\partial R_i/\partial \beta$]
{\label{zqder} \small
Fit results for $\nu$ from fitting $\partial R_i/\partial \beta$,
improved action, with eq.~(\ref{fitQp}).  The data with better
scaling behaviour belong to $\partial R_2/\partial \beta$.
}}
\end{center}
\end{figure}

We fitted our data for the derivatives of the $R_i$ with respect to
$\beta$, according to eq.~(\ref{fitQp}).  The results are given in
table~\ref{Qp}.  The $\nu$--estimates of this table are plotted in
figure~\ref{zqder}.

Obviously, the derivatives of the cumulant scale better than those of
$Z_a/Z_p$: while the cumulant's derivative gives a small $\chi^2$/dof
already for $L_{\rm min}=6$, for the $R_1$'s derivative one needs
$L_{\rm min}=18$ in order to have a small $\chi^2$/dof and to reach
stability of the result.  However, also in this case the leading
correction to scaling is strongly suppressed.  Note that the range of
lattice sizes considered here is relatively small: the difference in
scaling behaviour is actually due to a bigger amplitude of the
next-to-leading correction to scaling, as discussed below.

The systematic error estimates due to neglecting leading order
correction to scaling are given, as usual, in square brackets.  Their
evaluation followed the procedure used in section~5.2.2.  We defined
the tilde quantities by
\be
\tilde \frac{\partial R_i}{\partial \beta}
= \frac{\partial R_i}{\partial \beta} - a_i \, 
\frac{b^{(S)}_i}{22} \, L^{1/\nu-0.81}   \, , 
\ee
where $a_i$ are given in table~\ref{Qp} and and the leading correction
amplitudes $b^{(S)}_i$ of the standard action are taken from
table~\ref{BQder2} and table~\ref{BZder2}.  The absolute difference of
the $\nu$ obtained fitting to eq.~(\ref{fitQp}) the tilde and the
original Monte Carlo data are the error estimates.

\BT{r|l|l|c}
\mc{4}{c}{\bf Improved Action}\\[2mm]
$ L_{\rm min}$ & \phantom{xx} $a_1$ & \phantom{xx} $\nu$& $\chi^2$/dof \\
\hline
10&--1.1419(12) & 0.62850(14)[32]\{152\}& 3.08\\
16&--1.1487(23) & 0.62924(25)[27]\{74\}& 0.90\\
20&--1.1543(39) & 0.62979(39)[26]\{43\}& 0.32\\
24&--1.1552(51) & 0.62988(51)[32]\{51\}*& 0.40\\
\hline
\mc{4}{c}{ }\\
$L_{\rm min}$ & \phantom{xx} $a_2$  & \phantom{xx} $\nu$  &   $\chi^2$/dof
\\
 \hline
 6& 0.66160(42)&   0.62969(11)[22]\{41\} &  0.79 \\
 8& 0.66247(63)&   0.62987(14)[18]\{26\} &  0.57 \\
10& 0.6622(11) & 0.62982(22)[16]\{15\}*  &  0.60 \\
12& 0.6626(12) & 0.62989(24)[24]\{11\}   &  0.64 \\
 \hline
\end{tabular}
\parbox[t]{.85\textwidth}
 {
 \caption[IA, fit $\partial R_i/\partial \beta$, no correction to scaling]
 {\label{Qp}
Fit of $\partial R_1/\partial \beta$ (top) and $\partial R_2/\partial
 \beta$ (bottom) with eq.~(\ref{fitQp}).
}}
\ET

The systematic error estimates due to subleading corrections are given in
curly brackets.
To  estimate them, we used a fit  ansatz
\be\label{Rder}
\frac{\partial R_i}{\partial \beta}
= a_i \, L^{1/\nu} ( 1 + b_i \, L^{-x} ) \, ,
\ee
and then defined as usual the tilde quantities subtracting 
$a_i b_i  L^{1/\nu-2}$ from the Monte Carlo data.
The amplitudes $b_i$ are given in table~\ref{Adz3}.
Comparing the fit without correction to scaling of the Monte Carlo and of 
the tilde data we obtained  the estimates given in the curly brackets.

\BT{l|c|c|c|c}
\mc{5}{c}{\bf Improved Action}\\[2mm]
\phantom{x} $x$ &  $a_1$ & $\nu$ & $b_1$&$\chi^2$/dof\\
\hline
0.81&--1.20(56)\mxx&0.63340(46)&--0.163(12)&0.68\\
2.4\mx&--1.1561(11)&0.62993(14)&--1.309(31)&0.34\\
2.33(28)  &--1.1572(41)&0.63003(40)&--1.18(68)\mx &0.36\\
\hline
\mc{5}{c}{ } \\
\phantom{x} $x$ &  $a_2$ & $\nu$ & $b_2$&$\chi^2$/dof\\
\hline
2.0\mx&\mx0.6637(12)&0.63009(25)&--0.058(29)&0.63\\
2.4\mx&\mx0.6635(11)&0.63005(23)&--0.103(51)&0.62\\
2.8\mx&\mx0.6633(10)&0.63002(21)&--0.190(93)&0.61\\
\hline
\end{tabular}
\PB
 {
\caption[IA, fit $\partial R_i/\partial \beta$, with correction to scaling
]
{\label{Adz3} \small
Fitting $\partial R_1/\partial \beta$ (top)
and $\partial R_2/\partial \beta$ (bottom)
with eq.~(\ref{Rder}).
}}
\end{center}
\end{table}

\begin{figure}
\begin{center}
\includegraphics[width=10cm]{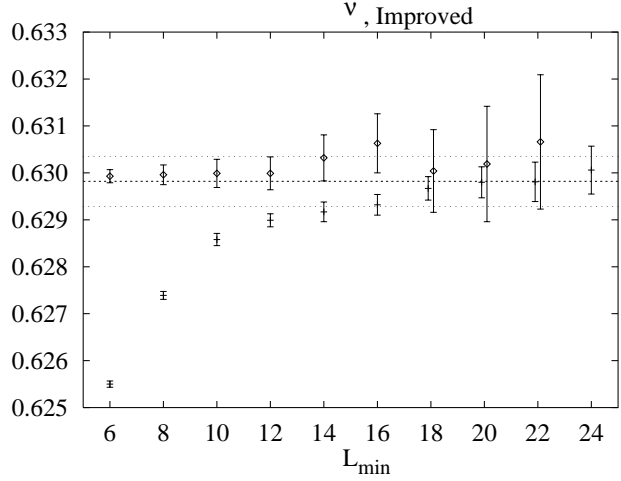}
\parbox[t]{.85\textwidth}
 {
 \caption[IA, $\nu$ from $\partial R_1/\partial \beta$, without and with
 correction to scaling]
 {\label{figZder}
\small
Fits of $\partial R_1/\partial \beta$, improved action, with
eq.~(\ref{fitQp}) (bars)
and with eq.~(\ref{Rder}) and $x=2.4$ fixed (diamonds).
 }}
\end{center}
\end{figure}

\begin{figure}
\begin{center}
\includegraphics[width=10cm]{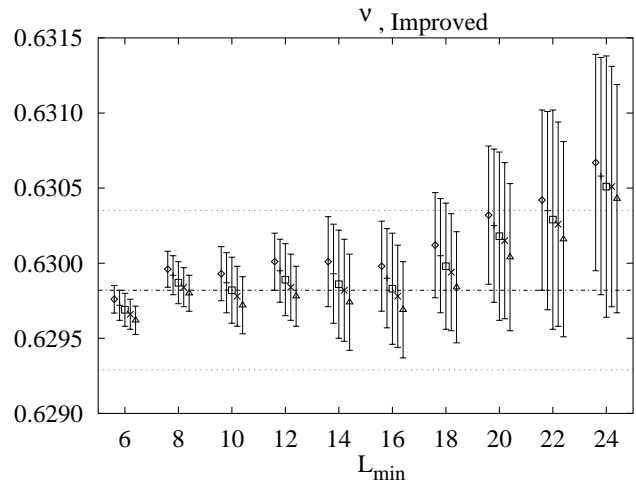}
\parbox[t]{.85\textwidth}
 {
 \caption[IA, $\nu$ from $\partial R_2/\partial \beta$,
 $\beta$-dependence]
{\label{betanu} \small
  Fit results for $\nu$ from fitting $\partial R_2/\partial \beta$,
  improved action, with eq.~(\ref{fitQp}).  The $\nu$-estimate is
given for the five $\beta$-values 0.383243 to 0.383247 in steps of
 0.000001 (left to right).
 }}
\end{center}
\end{figure}

Let us make some comments on the results given in table~\ref{Adz3}.
Including very small lattice sizes, the $R_1$ derivative's deviations
from eq.~(\ref{fitQp}) are strong enough to allow a fit with a free
effective exponent, which turns out to be $2.33(28)$, i.e.\ 
again of order 2.  Notice again that enforcing $x=0.81$ does not
prevent the fit from giving an acceptable $\chi^2$.  We have to rely
on the universality of the ratios of correction amplitudes, as
discussed before, to rule out this fit.

In figure~\ref{figZder} we plot the fits to $\partial R_1 /\partial
\beta$ using eq.~(\ref{Rder}) for various values of $L_{\rm min}$ with
fixed exponent $x=2.4$, together with the fits without correction
to scaling of table~\ref{Qp}.  This is sufficient to demonstrate the
nice scaling behaviour of the derivative of $R_1$ once an effective
next-to-leading correction to scaling exponent of order 2 has been
included in the fit.

In the case of $\partial R_2 /\partial \beta$ the corrections are too
small to allow for a free effective exponent fit.  Therefore we fix
the $x$-values between 2.0 and 2.8, as suggested by the fits on
$\partial R_1 /\partial \beta$.  The $\nu$-results are nicely stable
within these bounds and consistent with the values obtained without
scaling corrections.

These fits show clearly that the better scaling behaviour of the
Binder cumulant derivative is due to a smaller amplitude of the
next-to-leading corrections to scaling compared to that of the
$Z_a/Z_p$ derivative.

Finally, we checked also for the systematic dependence on the location
of $\beta_c$ for $R'_2$, which is giving the nicest result.  We
repeated the fits for the $Q$-derivative on data from five shifted
$\beta$-values ranging from 0.383243 to 0.383247 in steps of 0.000001,
covering thus two standard deviations around our $\beta_c$ estimate.
The results for $\nu$ are shown in figure~ \ref{betanu}.  The effects
of this variation is neglible compared with the errors of
table~\ref{Qp}.

We thus quote as our final estimate for $\nu$
\begin{center}
\begin{tabular}{rrll}
{\bf improved action:}  &$\nu$ &=~ 0.62988(51)[32]\{51\} 
                       &from $\partial R_1/\partial \beta$  \\[1mm]
                       &$\nu$ &=~ 0.62982(22)[16]\{15\} 
                       &from  $\partial R_2/\partial \beta$
\end{tabular}
\end{center}
The final estimate of $\nu$ appears with dotted error-lines in
figures~\ref{zqder}, \ref{betanu} and \ref{figZder}. Recall 
that to this end the statistical and systematic errors were 
added up.

\subsection{Fitting the Susceptibility}

\subsubsection{Standard Action, Fit $\chi$}
\BT{c|c|c|c|c|c}
\mc{6}{c}{\bf Standard Action}\\[2mm]
$L_{\rm min}$&$ c$ & $ d$&$\eta$ & $f$&$\chi^2$/dof \\
 \hline
10& --0.515(83)&1.5600(44)& 0.03751(62)& --0.585(12)&0.96\\%
12& --0.67(24)\mx&1.5548(84) & 0.0368(11)\mx& --0.568(27)&0.98\\
14& --1.00(41)\mx&1.546(12)\mx&0.0358(15)\mx&--0.537(41)& 1.06\\
 \hline
 \end{tabular}
\PB
 {
 \caption[SA, fit $\chi$ at fixed $Q$]
{\label{Bchi1} \small
 Fit of $\chi$ at fixed $Q=1/1.604=0.62344$, 
using eq.~(\ref{fixiomega}) with $\omega=0.81$.
}}
\ET

\BT{c|c|c|c|c|c}
\mc{6}{c}{\bf Standard Action}\\[2mm]
$L_{\rm min}$ & $ c$ & $ d$&$\eta$ & $f$&$\chi^2$/dof \\
 \hline
  8&--1.08(8)\mx  &  1.543(4)\mx & 0.0354(5)\mx &--0.096(12) & 1.22 \\
 10&--0.78(14)    &  1.553(6)\mx & 0.0366(8)\mx &--0.130(19) & 1.14\\
 12&--1.2(6)\mx\mx   & 1.544(16)& 0.0355(19)    &--0.095(61) &1.15\\
 14&--1.5(8)\mx\mx   & 1.538(15)& 0.0348(19)    &--0.069(68) & 1.24\\
\hline
\end{tabular}
\PB
{
\caption[SA, fit $\chi$ at fixed $Z_a/Z_p$]
{\label{Zchi1} \small
 Fit of $\chi$ at fixed $Z_a/Z_p=0.5425$,
 using eq.~(\ref{fixiomega}) with $\omega=0.81$.
}}
\ET
It turns out that the estimate of $\eta$ from fits of the magnetic
susceptibility $\chi$ taken at $\beta_c$ depends quite strongly on the
value of $\beta_c$. Taking the magnetic susceptibility at a fixed
value of a phenomenological coupling removes this problem, as discussed
in ref.~\cite{ballold}.  One defines a function $\beta(L)$ by
requiring that for any $L$ the relation $R_i(L,\beta(L))= const $
holds. 
The susceptibility is then computed at $\beta(L)$. 
We performed this analysis for the two cases of fixing $Q=0.6240$ and
fixing $Z_a/Z_p =0.5425$.  Note that in principle any value for
$Q$ and $Z_a/Z_p$ that can be taken by the phenomenological
couplings would work. However, for practical purposes it is the best to take
good approximations of $R_i^*$.

We fitted our data to the ansatz: 
\be\label{fixiomega}
\chi~\left(L, \beta_c(L)\right) = c +   d \,  L^{2-\eta} \, 
 \left( 1 +  f \, L^{-\omega}  \right) \, , 
\ee
where $c$ is the leading analytic part of $\chi$, and $f \,
L^{-\omega}$ gives leading order corrections.  The results of the fits
are given in table~\ref{Bchi1} and in table~\ref{Zchi1}.  In both
cases an acceptable $\chi^2/$dof is reached at $L_{\rm min}=10$.  It
is interesting to see that the correction to scaling amplitude $f$ is
considerably smaller in the case of fixed $Z_a/Z_p$. Therefore it
seems reasonable to assume that also $L^{-2 \omega}$ corrections are
smaller for fixed $Z_a/Z_p$.  We thus take $\eta$ from fixed $Z_a/Z_p$
at $L_{\rm min}=10$ as our final result. As estimate of the systematic
error we quote the difference to the fixed $Q$ result at $L_{\rm  min}=10$:
$$
\mbox{\bf standard action: ~~ }  \eta = 0.0366(8)\{9\}  \, .
$$
We have checked that the uncertainty in the estimate of $\omega$ leads
to negligible errors in $\eta$. We also performed fits of the magnetic
susceptibility without a constant term in the ansatz.  It is
reassuring that the results for $\eta$ are consistent with those found
above, when $L_{\rm min}=20$ is taken.

\subsubsection{Improved Action, Fit $\chi$}

\BT{r|l|l|l|c}
\mc{5}{c}{\bf Improved Action}\\[2mm]
$ L_{\rm min}$ 
& \phantom{xx} $c$ & \phantom{xx} $d$ & \phantom{xx}$\eta$ 
&  $\chi^2$/dof \\
 \hline
4 &--0.459(8)  & 0.9496(9) & 0.0351(3)[17] & 0.42 \\
6 &--0.540(56) & 0.9516(18)& 0.0357(6)[14] & 0.28 \\
8 &--0.553(77) & 0.9519(20)& 0.0358(6)[14] & 0.30 \\
10&--0.58(13)  & 0.9525(29)& 0.0359(9)[12] & 0.32 \\
12&--0.55(13)  & 0.9519(28)& 0.0358(8)[12] & 0.33 \\
\hline
\end{tabular}
\PB
{
\caption[IA, fit $\chi$ at fixed $Q$]
{\label{ccc2}
\small 
Fitting $\chi$ with eq.~(\ref{fixi}) at fixed $Q=0.6240$. 
}}
\ET

\BT{r|l|l|l|c}
\mc{5}{c}{\bf Improved Action}\\[2mm]
$ L_{\rm min}$ 
& \phantom{xx} $ c$ & \phantom{xx} $ d$ & \phantom{xx}$\eta$ 
&  $\chi^2$/dof \\
 \hline
 4 &--0.541(49) & 0.9545(17)& 0.03664(53)[20]*& 0.15 \\
 6 &--0.538(61) & 0.9544(20)& 0.03662(64)[20] & 0.15 \\
 8 &--0.532(62) & 0.9543(20)& 0.03657(60)[20] & 0.15 \\
 10&--0.511(90) & 0.9538(24)& 0.03644(75)[18] & 0.15 \\
 12&--0.53(16)  & 0.9541(32)& 0.03651(94)[16] & 0.16 \\
\hline
\end{tabular}
\PB
 {
\caption[IA, fit $\chi$ at fixed $Z_a/Z_p$]
 {\label{ccc3}
 \small 
 Fitting $\chi$ with eq.~(\ref{fixi}) at fixed $Z_a/Z_p=0.5425$. 
 }}
\ET

\begin{figure}
\begin{center}
{\includegraphics[width=10cm]{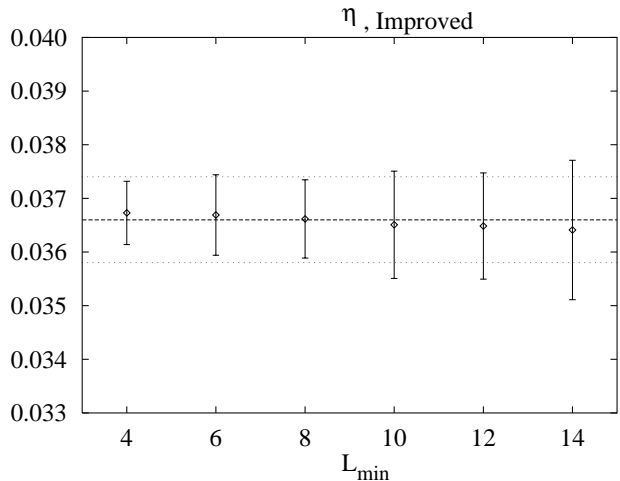}}
\PB
 {
 \caption[IA, $\eta$ from $\chi$ at fixed $Z_a/Z_p$]
 {\label{aa1}
\small 
$\eta$-estimates, fitting 
$\chi$-data at fixed $Z_a/Z_p$ to eq.~(\ref{fixi}), improved action.
}}
\end{center}
\end{figure}

Also in the case of the improved action we computed the magnetic 
susceptibility at fixed $Q=0.6240$ and at fixed $Z_a/Z_p=0.5425$. 
We fitted
our data with the ansatz
\be\label{fixi}
\chi~(L, \beta(L)) =  c +  d  \,  L^{2-\eta} \, . 
\ee
Here we have skipped the term $ L^{-\omega}$.
Our fit results are given in table~\ref{ccc2} and plotted 
in figure~\ref{aa1}.

{}From the Ising model with the standard action we know that 
the amplitude of the $L^{-\omega}$ correction is much smaller
for $\chi$ at fixed $Z_a/Z_p $ than at fixed $Q$ (see below).
Therefore the 
comparison of both results gives a nice check of the systematic
errors introduced by the ommision of a $L^{-\omega}$ term in the fit 
ansatz. 
Since the corrections are smaller in the case of fixed $Z_a/Z_p$,
we quote the corresponding result as our final estimate. 

The systematic error due to the ommision of a $L^{-\omega}$ term
in the ansatz is computed in the same way as in the previous sections.
Namely, we defined the tilde quantities as 
\be
\tilde \chi(L) = \chi(L) -
d \, \frac{f^{(S)}}{22} \, L^{2-\eta-0.81}   \, , 
\ee
where $f^{(S)}$ are taken from table~\ref{Bchi1} for $\chi$ at fixed $Q$ 
and table~\ref{Zchi1} for $\chi$ at fixed $Z_a/Z_p$.
Then we compare the results obtained fitting the $\tilde \chi$ with 
eq.~(\ref{fixi}) and the $\chi$, both at fixed $Q$ and fixed $Z_a/Z_p$.
The absolute differences of the $\eta$ obtained in this way are the estimates
of the systematic errors. 
Notice that in this case the next-to-leading correction to scaling enters with
an exponent of order 2 which is somehow already taken into account with the 
analytical contribution denoted by $c$ in our fit ans\"atze. 
Therefore we only quote the systematic error due to the leading correction.
 
We haven choosen the result of $L_{\rm min}=8$ as our final estimate, since 
it is consistent with the result obtained from
$L_{\rm min}=4$ and $L_{\rm min}=6$.
Hence our final estimate is 
$$
\mbox{\bf improved action: ~~ }  \eta =0.0366(6)[2]\, .
$$
which is consistent with the estimate obtained from the standard action.

\subsection{Fitting the Energy}
As a consistency check we tried to obtain the exponent $\nu$ from
the singular behaviour of the energy. Since it turned out that the 
statistical errors of the results for $\nu$ are larger than those 
obtained from
the derivatives of the $R_i$ we skipped the elaborate analysis of 
systematic errors that was used above.

\subsubsection{Improved Action, Fit $E$}

The expectation values of the energy $E$ in  the spin-1 model, 
defined through eq.~(\ref{EEE}) where subjected to a fit with 
\be\label{enerfit1}
E = a_E + b_E \, L^{\, 1/\nu-3} \, . 
\ee  
The results are quoted in table~\ref{Ae2}.
The dependence of the estimate for $\nu$ on $L_{\rm min}$ is shown
in figure~\ref{figEn}, where the estimates obtained from the standard 
action are also given. 

The results are consistent with the previous estimate of $\nu$ but
with bigger errors.  The final $\nu$-estimate, with
error-lines, obtained from the cumulant derivative for the improved
action is plotted for comparison.

\subsubsection{Standard Action, Fit $E$}

We first tried a fit without corrections, using eq.~(\ref{enerfit1}).
The fit results for the nearest neighbour energy $E_{NN}$ of the
standard action, defined in eq.~(\ref{ENN}), are given in
table~\ref{Be1}.

\BT{r|c|c|c|c}
\mc{5}{c}{\bf Improved Action}\\[2mm]
$L_{\rm min}$ & $a_E$& $b_E$ & $\nu$& $\chi^2$/dof \\
 \hline
 6& 1.23156(14)&--0.9884(18)   &0.62893(39)  &0.58\\
10& 1.23155(23)&--0.9929(64)   &0.6297(11)\mx&0.63\\
14& 1.23156(29)&--0.987(11)\mx &0.6288(18)\mx&0.67\\
18& 1.23157(47)&--0.983(21)\mx &0.6281(32)\mx&0.77\\
 \hline
 \end{tabular}
\PB
 {
 \caption[IA, fit energy, no correction to scaling]
{\label{Ae2} \small Fit of the  energy  $E$ with eq.~(\ref{enerfit1}).
}}
\ET

\BT{c|c|c|c|c}
\mc{5}{c}{\bf Standard Action}\\[2mm]
$ L_{\rm min}$ &$a_E$ &$b_E$&$\nu$&$\chi^2$/dof\\
\hline
16&0.3302003(32)& 0.7331(18)&0.62902(36)&0.73\\
20&0.3302028(47)& 0.7349(39)&0.62937(72)&0.75\\
28&0.3302008(65)& 0.7325(74)&0.6290(13)\mx&0.81\\
36&0.330201(12)\mx& 0.733(19)\mx&0.6290(30)\mx&0.97\\
\hline
\end{tabular}
\PB
 {
 \caption[SA, fit energy, no correction to scaling] 
{\label{Be1}
\small
Fit of the energy $E_{NN}$ with eq.~(\ref{enerfit1}).
}}
\ET

\begin{figure}
\begin{center}
\includegraphics[width=10cm]{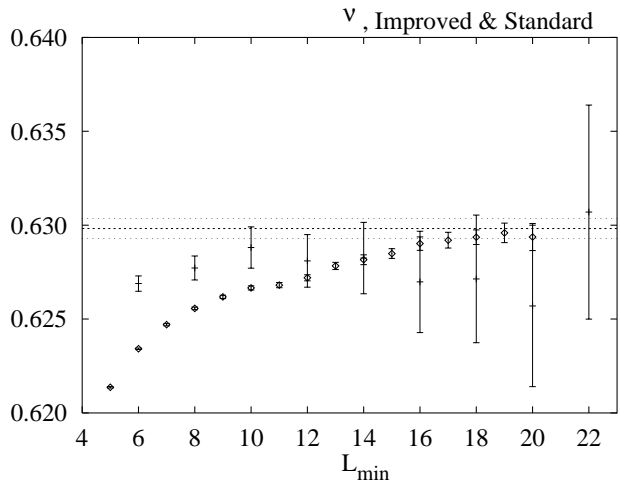}
\PB
 {
 \caption[IA \& SA, $\nu$ from energy]
 {\label{figEn}
\small
Fits of the energy with eq.~(\ref{enerfit1}) for both actions
(without correction to scaling).
}}
\end{center}
\end{figure}

Compared to the case of the improved action, the table starts with
much larger $L_{\rm min}$. The corrections to scaling are much bigger
than in the improved case. This becomes obvious from a look at
figure~\ref{figEn}, which shows the dependence of the $\nu$-estimate
on $L_{\rm min}$.  Therefore we take into account corrections
to scaling with the ansatz

\be\label{enerfit2}
E_{NN} = a_E + b_E \, L^{\, 1/\nu-3} \left( 1 + c_E L^{-\omega} \right) \, . 
\ee  
The results given in table~\ref{Be2} for $\omega=0.81$ show a nice agreement, 
however with large errors. 

\BT{r|c|c|c|c|c}
\mc{6}{c}{\bf Standard Action}\\[2mm]
$ L_{\rm min}$ &$a_E$ &$b_E$&$\nu$&$c_E$&$\chi^2$/dof\\
 \hline
8 &0.3302162(32)&0.7693(35)   &0.63372(53)  &--0.155(10)& 1.11\\
14&0.3302057(59)&0.754(12)\mx &0.6317(16)\mx&--0.098(44)& 0.87\\
18&0.3302093(93)&0.756(30)\mx &0.6321(39)\mx&--0.09(14)\mx& 0.85\\
 \hline
 \end{tabular}
\parbox[t]{.85\textwidth}
 {
 \caption[SA, fit energy, with correction to scaling]
{\label{Be2} \small Fit of the energy $E_{NN}$ with
eq.~(\ref{enerfit2}) and $\omega=0.81$.
}}
\ET

\section{Comparison with Other Estimates}

It is interesting to compare our estimates with other ones available
from the literature.  Note that extensive tables with many data from
literature and experiment can be found in refs.~\cite{bloete},
\cite{guida}, and \cite{butera}.

Some of the more recent estimates for the $\nu$ and $\eta$ exponents,
together with the critical coupling of the standard Ising model are
collected in table \ref{exp1}.  Our estimates are given in the last
two lines: the upper one gives to the estimates obtained with the
standard action (SA) while the lower one those from the improved
spin--1 action (IA). The underlined estimates are obtained by 
choosing our best estimate from the improved action and adding 
the statistical and systematic error estimates in order 
to obtain the overall uncertainty. 

The value we obtained for the Binder cumulant from the improved action
$Q=0.62393(13)[35]\{5\}$ can be compared with the estimate
$Q=0.6233(4)$ of ref.~\cite{bloete}.  It is reassuring that the high
precision estimates of the recent years seem to be nicely consistent
with each other.

\BT{c|c|l|l|l}
Ref. & Method & \phantom{xxx}$\nu$ &\phantom{xx} $\eta$ &
\phantom{xxxx}$\beta_c$\\
 \hline
\cite{guida}     & EPS   & 0.6293(26)  & 0.036(6)    &\\   % guida free bc
\cite{guida}     & 3D FT & 0.6304(13)  & 0.0335(25)  &\\
\cite{nickel}    & 3D FT & 0.6301(5)   & 0.0355(9)   &\\   % cited by guida
\cite{butera}    & HT    & 0.6310(5)   &             &\\   % gamma = 1.2385(5)
\cite{hapi}      & MC    & 0.6308(10)  &             &\\
\cite{bloete,talapov}&MC & 0.6301(8)   & 0.037(3)      & 0.2216544(3)[3]\\ 
\cite{ball1}     & MC    & 0.6294(5)[5] & 0.0374(6)[6] & 0.22165456(15)[5]\\ 
\hline
SA & MC, $R'_1$  & 0.63014(54)\{55\}  &  & 0.22165431(19)\{15\} \\
SA & MC, $R'_2$  & 0.62973(43)\{46\}$\LL 14\GG$  &  & 0.22165405(23)\{25\} \\
SA & MC, $\chi$  & & 0.0366(8)\{9\}  &  \\
IA & MC, $R'_2$  & 0.62982(22)[16]\{15\}  &  & \\
IA & MC, $R'_2$  & \underline{0.6298(5)}              &  & \\
IA & MC, $R'_1$  & 0.62988(51)[32]\{51\}  &   & \\
IA & MC, $\chi$  &                       & 0.0366(6)[2]  & \\
IA & MC, $\chi$  &                       & \underline{0.0366(8)}  & \\[1mm]
\hline
\end{tabular}
\PB
{
\caption[Comparing Exponent Estimates with Previous Ones]
{\label{exp1}
Results of the present study from standard (SA) and improved (IA) actions 
are compared with other estimates: from $\epsilon$--expansion (EPS),
field theory calculations in three dimensions (3D FT), high temperature 
expansions (HT) and Monte Carlo simulations (MC). The underlined estimates 
for the critical exponents are our best estimates together with 
error estimates which give the overall uncertainty, including
systematic effects. 
}}
\ET

\section{Conclusions}

By performing a detailed comparison with high precision results of the
standard action, we have demonstrated that the spin-1 Ising model with
suitably tuned coupling constants has remarkably improved finite size
scaling properties. We obtained estimates of very high precision for
the critical exponents $\nu$ and $\eta$ and two other universal
quantities, the Binder cumulant $Q$ and the ratio of partition
functions $Z_a/Z_p$. 

All estimates from the two different actions are
consistent with each other. In spite of the higher
statistics and the bigger lattice sizes of the standard action data,
the estimates from the improved action are by far more precise.
In particular, c.f.\ table~\ref{exp1}, the systematic errors
are smaller for the improved action than for the standard action.

The authors of refs.~\cite{ball1,ball2} claim that an improvement of
the action as discussed in this paper and in ref.~\cite{ball2} does
not allow for more precise estimates of universal quantities such as
the critical exponents.  In their argument they ignore the fact that
ratios of correction amplitudes are universal. Once these ratios are
computed for the standard Ising model, where the corrections are
large, they allow for powerful bounds in the case of the improved
action.  E.g., leading order corrections to scaling of the Binder
cumulant are much stronger than those of the derivative of the Binder
cumulant.  Therefore it is quite clear, that we can safely ignore
$L^{-\omega}$ corrections in the analysis of $ Q'$ obtain from the
improved model.

It would be worthwhile to use the present model in studies of physical
quantities not discussed in this work and to check to what extent the
improved scaling behaviour helps to get better estimates.

An interesting question is whether and to what extent the improvement
that we have achieved can be further enhanced. It seems tempting to
study models with even more couplings in order to systematically
reduce the effects from next-to-leading corrections to scaling.
However, little is known about these higher order corrections, and one
does not really know how many parameters would be necessary to remove
corrections of order $L^{-x}$, with $x \approx 2$.  Note that at that
level of improvement also the question of improved observables comes
into play.  We thus believe that the improvement with two parameters
which we performed is the optimal thing one can do in a systematic
way.  Note that our improvement also eliminates subleading corrections
of the typ $L^{-n\omega}$, cf.\ the discussion in section~5.

Last but not least, application of the ideas underlying the present
analysis to other models seems very promising.

%%%%%%%%%%%%%%%%%%%%%%%%%%%%%%%%%%%%%%%%%%%%%%%%%%%%%%%%%%%%%%%%%%%%%%%%%%%%%
% Appendix 

\BT{r|l|l|l|c}
\mc{5}{c}{\Large \bf  Appendix: Basic Monte Carlo Results} \\[2cm]
\mc{5}{c}{\bf Standard Action} \\[2mm]
$L$ & \phantom{xx} $Z_a/Z_p$ & \phantom{xx} $Q$ & \phantom{xx} $\chi/L^2$ 
& $E_{NN}$ \\
\hline
  4 &              &0.6597860(58) &1.324852(21) &0.4310732(41) \\
  5 &              &0.6537418(59) &1.350645(22) &0.4047035(33) \\
  6 & 0.550587(31) &0.6492132(60) &1.363074(23) &0.3881589(27) \\
  7 &              &0.6458092(62) &1.369365(24) &0.3769937(24) \\
  8 & 0.549195(34) &0.6431935(72) &1.372565(28) &0.3690452(23) \\
  9 &              &0.6411374(85) &1.374004(34) &0.3631459(23) \\
 10 & 0.548166(50) &0.6394995(96) &1.374541(37) &0.3586312(22) \\
 11 &              &0.6381157(99) &1.374188(38) &0.3550664(20) \\
 12 & 0.547379(52) &0.636980(11)  &1.373593(41) &0.3522077(19) \\
 13 &              &0.636038(15)  &1.372799(58) &0.3498712(24) \\
 14 & 0.546851(71) &0.635224(17)  &1.371762(63) &0.3479251(23) \\
 15 &              &0.634494(18)  &1.370545(66) &0.3462842(22) \\
 16 & 0.546212(76) &0.633875(19)  &1.369375(69) &0.3448912(21) \\
 17 &              &0.633329(20)  &1.368077(72) &0.3436884(20) \\
 18 &              &0.632816(23)  &1.366696(80) &0.3426440(20) \\
 19 &              &0.632386(24)  &1.365440(83) &0.3417330(19) \\
 20 & 0.54587(12)  &0.631947(89)  &1.36379(31)  &0.3409214(70) \\
 24 & 0.54552(12)  &0.630659(82)  &1.35886(28)  &0.3384943(49) \\
 28 & 0.54512(13)  &0.629767(93)  &1.35386(32)  &0.3368733(46) \\
 32 & 0.54499(12)  &0.62906(14)   &1.34917(40)  &0.3357286(49) \\
 40 & 0.54447(21)  &0.62805(15)   &1.34171(52)  &0.3342390(47) \\
 48 & 0.54432(20)  &0.62720(17)   &1.33415(58)  &0.3333177(42) \\
 56 & 0.54383(32)  &0.62705(23)   &1.32848(81)  &0.3327115(48) \\
 64 & 0.54354(30)  &0.62690(23)   &1.32378(82)  &0.3322842(41) \\
 80 & 0.54373(51)  &0.62568(34)   &1.3144(13)   &0.3317204(49) \\
 96 & 0.54263(49)  &0.62585(36)   &1.3079(13)   &0.3313780(38) \\
112 & 0.54546(83)  &0.62412(64)   &1.2969(20)   &0.3311406(48) \\
128 & 0.54337(80)  &0.62518(62)   &1.2921(22)   &0.3309828(45) \\
\hline
 \end{tabular}
\PB
 {
 \caption[SA, Monte Carlo estimates for $R_1$, $R_2$, $\chi$, and $E_{NN}$]
 {\label{ZQXstandard}
\small
Results for $R_1=Z_a/Z_p$, $R_2=Q$, the susceptibility $\chi$ 
divided by $L^2$, and the energy per link, $E_{NN}$, 
at $\beta= 0.2216545$. 
}}
\ET

\BT{r|c|c|l}
\mc{4}{c}{\bf Improved action} \\[2mm]
$L$ & $Z_a/Z_p$ & $Q$ & \phantom{x} $\chi/L^2$ \\
\hline
 4& 0.53599(07)&0.62408(05)& 0.87609(13) \\
 6& 0.54080(07)&0.62491(06)& 0.88104(14) \\
 8& 0.54193(09)&0.62460(07)& 0.87706(16) \\
10& 0.54209(18)&0.62464(13)& 0.87272(33) \\
12& 0.54248(19)&0.62416(14)& 0.86778(35) \\
14& 0.54230(21)&0.62426(15)& 0.86399(38) \\
16& 0.54263(17)&0.62409(13)& 0.86000(32) \\
18& 0.54251(23)&0.62410(17)& 0.85697(42) \\
20& 0.54286(28)&0.62379(21)& 0.85332(50) \\
22& 0.54244(30)&0.62404(23)& 0.85115(55) \\
24& 0.54229(23)&0.62402(17)& 0.84914(43) \\
28& 0.54245(31)&0.62386(23)& 0.84373(55) \\
32& 0.54259(31)&0.62390(23)& 0.84010(56) \\
36& 0.54213(41)&0.62420(30)& 0.83742(73) \\
40& 0.54227(32)&0.62403(23)& 0.83392(56) \\
48& 0.54250(39)&0.62395(29)& 0.82803(69) \\
56& 0.54272(63)&0.62382(47)& 0.82351(12) \\
\hline
 \end{tabular}
\PB
{
 \caption[IA, Monte Carlo estimates for $R_1$, $R_2$ and $\chi$]
 {\label{ZQX}
\small
Results for $R_1=Z_a/Z_p$, $R_2=Q$, and the susceptibility $\chi$
divided by $L^2$, at $\beta= 0.383245$.
}}
\ET

\BT{r|c|c|c}
 \mc{4}{c}{\bf Improved Action} \\[2mm]
  $L$ & $E_{NN}$ &$ E_D$& $E$ \\
 \hline
 4 & 0.283630(29) & 0.647597(17) & 1.091901(43)\\
 6 & 0.248854(18) & 0.633033(10) & 1.152537(28)\\
 8 & 0.233532(15) & 0.626493(8)\CC & 1.178885(24)\\
10 & 0.225261(23) & 0.622948(12) & 1.193062(36)\\
12 & 0.220134(19) & 0.620744(9)\CC & 1.201829(31)\\
14 & 0.216748(18) & 0.619288(9)\CC & 1.207619(28)\\
16 & 0.214326(12) & 0.618243(6)\CC & 1.211750(19)\\
18 & 0.212558(14) & 0.617481(7)\CC & 1.214768(22)\\
20 & 0.211181(15) & 0.616886(7)\CC & 1.217115(24)\\
22 & 0.210132(14) & 0.616434(7)\CC & 1.218905(23)\\
24 & 0.209291(10) & 0.616073(5)\CC & 1.220344(16)\\
28 & 0.207992(10) & 0.615514(5)\CC & 1.222565(17)\\
32 & 0.207084(9)\CC & 0.615120(4)\CC & 1.224110(14)\\
36 & 0.206428(10) & 0.614838(5)\CC & 1.225227(16)\\
40 & 0.205910(7)\CC & 0.614613(3)\CC & 1.226109(11)\\          
48 & 0.205183(6)\CC & 0.614299(3)\CC & 1.227350(11)\\
56 & 0.204704(9)\CC & 0.614094(4)\CC & 1.228169(14)\\
 \hline
 \end{tabular}
\parbox[t]{.85\textwidth}
 {
 \caption[IA, Monte Carlo estimates for energies $E_{NN}$, $E_D$ and $E$]
 {\label{Ae1} \small 
  Results for energy $E_{NN}$, $E_D$ and $E$, at $\beta= 0.383245$.
}}
\ET                                       

\BT{r|l|l}
\mc{3}{c}{\bf Standard Action} \\[2mm]
$L$ & $ f(L)\, d R_1/d \beta $ 
& $f(L)\, d R_2/d \beta$ \\
\hline
   4 &               &    0.868988(29) \\
   5 &               &    0.866166(30) \\
   6 & --1.45953(13) &    0.863364(31) \\
   7 &               &    0.860978(32) \\
   8 & --1.47001(13) &    0.859086(37) \\
   9 &               &    0.857510(44) \\
  10 & --1.47311(18) &    0.856291(50) \\
  11 &               &    0.855176(52) \\
  12 & --1.47435(26) &    0.854265(56) \\
  13 &               &    0.853443(79) \\
  14 & --1.47492(26) &    0.852892(87) \\
  15 &               &    0.852278(93) \\
  16 & --1.47560(30) &    0.851801(97) \\
  17 &               &    0.85112(10)  \\
  18 &               &    0.85076(12)  \\
  19 &               &    0.85033(12)  \\
  20 & --1.47517(44) &    0.84968(52)  \\
  24 & --1.47462(45) &    0.84937(51)  \\
  28 & --1.47396(51) &    0.84789(58)  \\
  32 & --1.47305(64) &    0.84710(72)  \\
  40 & --1.47400(84) &    0.84641(94)  \\
  48 & --1.4737(14)  &    0.8462(10)   \\
  56 & --1.4714(13)  &    0.8457(14)   \\
  64 & --1.4756(16)  &    0.8454(15)   \\
  80 & --1.4749(20)  &    0.8479(21)   \\
  96 & --1.4721(49)  &    0.8451(24)   \\
 112 & --1.4716(37)  &    0.8451(41)   \\
 128 & --1.4727(53)  &    0.8433(41)   \\
\hline
 \end{tabular}
\PB
 {
 \caption[SA, Monte Carlo estimates for $d R_i/d\beta$]
 {\label{deriStan}
\small
Results for derivatives of $R_1=Z_a/Z_p$, $R_2=Q$, 
with respect to $\beta$, multiplied by the 
factor $f(L)=L^{-1/0.63}$, at $\beta= 0.2216545$. 
}}
\ET

\BT{r|l|l|l|l}
\mc{5}{c}{\bf Improved Action} \\[2mm]
$L$ & $ f(L)\,\partial R_1/\partial\beta $ 
& $f(L)\, d R_1/d \beta$  
& $f(L)\, \partial R_2/\partial\beta$ 
& $ f(L)\, d R_2/d \beta$  \\
\hline
 4&--1.09974(17)&--.63210(11)& .65894(15)&--.09882(03)\\
 6&--1.13595(17)&--.64829(10)& .66236(15)&--.09685(03)\\
 8&--1.14621(20)&--.65290(12)& .66295(18)&--.09614(03)\\
10&--1.15051(41)&--.65498(24)& .66276(37)&--.09577(07)\\
12&--1.15293(45)&--.65567(25)& .66307(40)&--.09571(07)\\
14&--1.15385(49)&--.65661(27)& .66310(46)&--.09555(08)\\
16&--1.15432(44)&--.65652(25)& .66290(39)&--.09554(07)\\
18&--1.15547(59)&--.65692(33)& .66294(55)&--.09545(11)\\
20&--1.15603(71)&--.65732(39)& .66322(64)&--.09553(11)\\
22&--1.15542(79)&--.65714(43)& .66310(74)&--.09574(17)\\
24&--1.15675(61)&--.65790(34)& .66419(55)&--.09568(11)\\
28&--1.15565(80)&--.65730(44)& .66199(75)&--.09520(20)\\
32&--1.15664(82)&--.65793(46)& .66283(75)&--.09523(18)\\
36&--1.1568(12) &--.65776(61)& .6638(11) &--.09525(49)\\
40&--1.15665(84)&--.65785(47)& .66326(84)&--.09557(37)\\
48&--1.1563(11) &--.65752(59)& .6635(11) &--.09513(58)\\
56&--1.1562(17) &--.65739(97)& .6628(19) &--.0958(12)\\
\hline
 \end{tabular}
\PB
 {
 \caption[IA, Monte Carlo estimates for $\partial R_i/\partial \beta$ and 
         $d R_i/d \beta$ ]
 {\label{deri}
Results for partial and total derivatives of $R_i$,
multiplied by the factor $f(L)=L^{-1/0.63}$, at $\beta= 0.383245$.
}}
\ET


\begin{thebibliography}{99}
\bibitem{nickel} Results by Murray and Nickel, taken from table 10 
                 of \cite{guida}. Errors from uncertainty of 
                 $\tilde g^*$ are not taken into account.
\end{thebibliography}
\end{document}